\documentclass[notitlepage,aps,prl,reprint,twocolumn,longbibliography,superscriptaddress]{revtex4-2}
\usepackage{graphicx}
\usepackage{amsmath}
\usepackage{amssymb}
\usepackage{comment}
\usepackage[colorlinks, allcolors=blue]{hyperref}

\usepackage[mathlines]{lineno}
\usepackage{braket}
\usepackage{wrapfig}
\usepackage{lipsum}
\usepackage{ulem}

\newcommand{\pref}[2]{\hyperref[#1]{\ref{#1}(#2)}}
\newcommand{\preff}[2]{\hyperref[#1]{\ref{#1 b}#2}}
\newcommand{\eqpref}[1]{\hyperref[#1]{(\ref{#1})}}

\newcommand{\squig}{{\raise.17ex\hbox{$\scriptstyle\sim$}}}

\begin{document}
\title{Observation of chiral solitary waves in a nonlinear Aharonov-Bohm ring}
\author{Ivan Velkovsky}
\thanks{These authors contributed equally to this work.}
\affiliation{Department of Physics, University of Illinois at Urbana-Champaign, Urbana, IL 61801-3080, USA}
\author{Anya Abraham}
\thanks{These authors contributed equally to this work.}
\affiliation{Department of Physics, University of Illinois at Urbana-Champaign, Urbana, IL 61801-3080, USA}
\author{Enrico Martello}
\thanks{These authors contributed equally to this work.}
\affiliation{School of Physics and Astronomy, University of Birmingham, Edgbaston, Birmingham B15 2TT, United Kingdom}
\affiliation{Dipartimento di Fisica e Astronomia ``Ettore Majorana", Universit\`{a} di Catania, I-95123 Catania, Italy}
\author{Jiarui Yu}
\affiliation{Department of Physics, University of Illinois at Urbana-Champaign, Urbana, IL 61801-3080, USA}
\author{Yaashnaa Singhal}
\affiliation{Department of Physics, University of Illinois at Urbana-Champaign, Urbana, IL 61801-3080, USA}
\author{Antonio Gonzalez}
\affiliation{Department of Physics, University of Illinois at Urbana-Champaign, Urbana, IL 61801-3080, USA}
\author{DaVonte Lewis}
\affiliation{Department of Physics, University of Illinois at Urbana-Champaign, Urbana, IL 61801-3080, USA}
\author{Hannah Price}
\email{H.Price.2@bham.ac.uk}
\affiliation{School of Physics and Astronomy, University of Birmingham, Edgbaston, Birmingham B15 2TT, United Kingdom}
\author{Tomoki Ozawa}
\email{tomoki.ozawa.d8@tohoku.ac.jp}
\affiliation{Advanced Institute for Materials Research (WPI-AIMR), Tohoku University, Sendai 980-8577, Japan}
\author{Bryce Gadway}
\email{bgadway@psu.edu}
\affiliation{Department of Physics, University of Illinois at Urbana-Champaign, Urbana, IL 61801-3080, USA}
\date{\today}
\affiliation{Department of Physics, The Pennsylvania State University, University Park, Pennsylvania 16802, USA}

\begin{abstract}
Nonlinearities can have a profound influence on the dynamics and equilibrium properties of discrete lattice systems.
The simple case of two coupled modes with self-nonlinearities gives rise to the rich bosonic Josephson effects.
In many-site arrays, nonlinearities yield a wealth of rich phenomena, including a variety of solitonic excitations, the emergence of vortex lattices in the presence of gauge fields, and the general support of chaotic dynamics.
Here, we experimentally explore a three-site mechanical ring with tunable gauge fields and nonlinearities. We observe a macroscopic self-trapping transition that is tunable by the magnetic flux, consistent with the equilibrium response. We further observe novel behavior that appears only out of equilibrium, the emergence of interaction-stabilized chiral solitary waves.
These results provide a starting point to explore nonlinear phenomena arising in larger mechanical arrays coupled to static and dynamical gauge fields.
\end{abstract}
\maketitle

Gauge fields play a fundamental role in our understanding of foundational topological phenomena such as the integer~\cite{vonKlitzing-IQHE} and fractional~\cite{Stormer-FQHE} quantum Hall effects.
Even in the classical (mean-field) limit, rich ground state structures result from the interplay of interactions and the kinetic frustration driven by gauge fields~\cite{Mueller-bosons-ladder,Jia-frustr}. The scientific richness of this problem has motivated a wide effort aimed at engineering artificial gauge fields in a range of experimental platforms, both quantum and classical, over the last two decades~\cite{AIDELSBURGER2018394}.

Considering the problem of classical interacting fields coupled to a magnetic flux~\cite{Mueller-bosons-ladder}, there will exist a wealth of dynamical phenomena distinct from the ground state response. Even for generic lattices, nonlinearities help to support, e.g., 
solitons and other stable self-trapped excitations~\cite{soliton-1,soliton-burnett,Self-trap-1,Soliton-Segev-Christo}.
While for the canonical two-mode problem of interacting, macroscopically occupied fields (i.e., the bosonic Josephson junction~\cite{Raghavan}), there exists a complete description of the rich nonlinear dynamics, the picture is less complete for many interacting modes. Even considering a ring of just three modes, the minimum needed to support a magnetic flux, the full characterization of dynamics is challenged by the added degrees of freedom, the breaking of integrability, and the concomitant emergence of chaos~\cite{Trimer-DST,Ham-Hopf-NEWMODE,Trimer0-munro,Trimer1-selftrapping-etc,Trimer2_Cao_2015,Trimer-selftrap-regimes,3-site-BHH,BHH-LeHur-synch}.
The dynamics of interacting flux rings is of growing recent interest due to the expanding capabilities for realizing nonlinear fluids coupled to gauge fields in atomic~\cite{Jia-frustr} and photonic~\cite{Mukh-top-soliton,NonlinearModesGaugeFields} systems as well as the potential connections to 
time-crystal-like phenomena based on interaction-induced chiral solitons~\cite{Wilczek-TC,Sacha-excited,Ohberg-gaugetheories,sacha-comment}.

Here, we explore the nonequilibrium dynamics of an elementary Josephson ring consisting of three mechanical oscillators with tunable self-nonlinearities and an artificial gauge field.
In the non-interacting regime, we observe chiral dynamics and a winding of the system's energy spectrum with applied flux. Adding nonlinearities, we observe a macroscopic self-trapping transition that is similarly tunable by the magnetic flux. We furthermore observe the emergence of self-localized chiral excitations, i.e., chiral solitary waves, for intermediate strengths of the nonlinearity.
These rich dynamics arising from the simple dissipation- and drive-free Aharonov-Bohm trimer suggest a wealth of possible dynamical phenomena achievable with active mechanics.

\begin{figure*}[t!]	                
    \includegraphics[width=1.85\columnwidth]
    {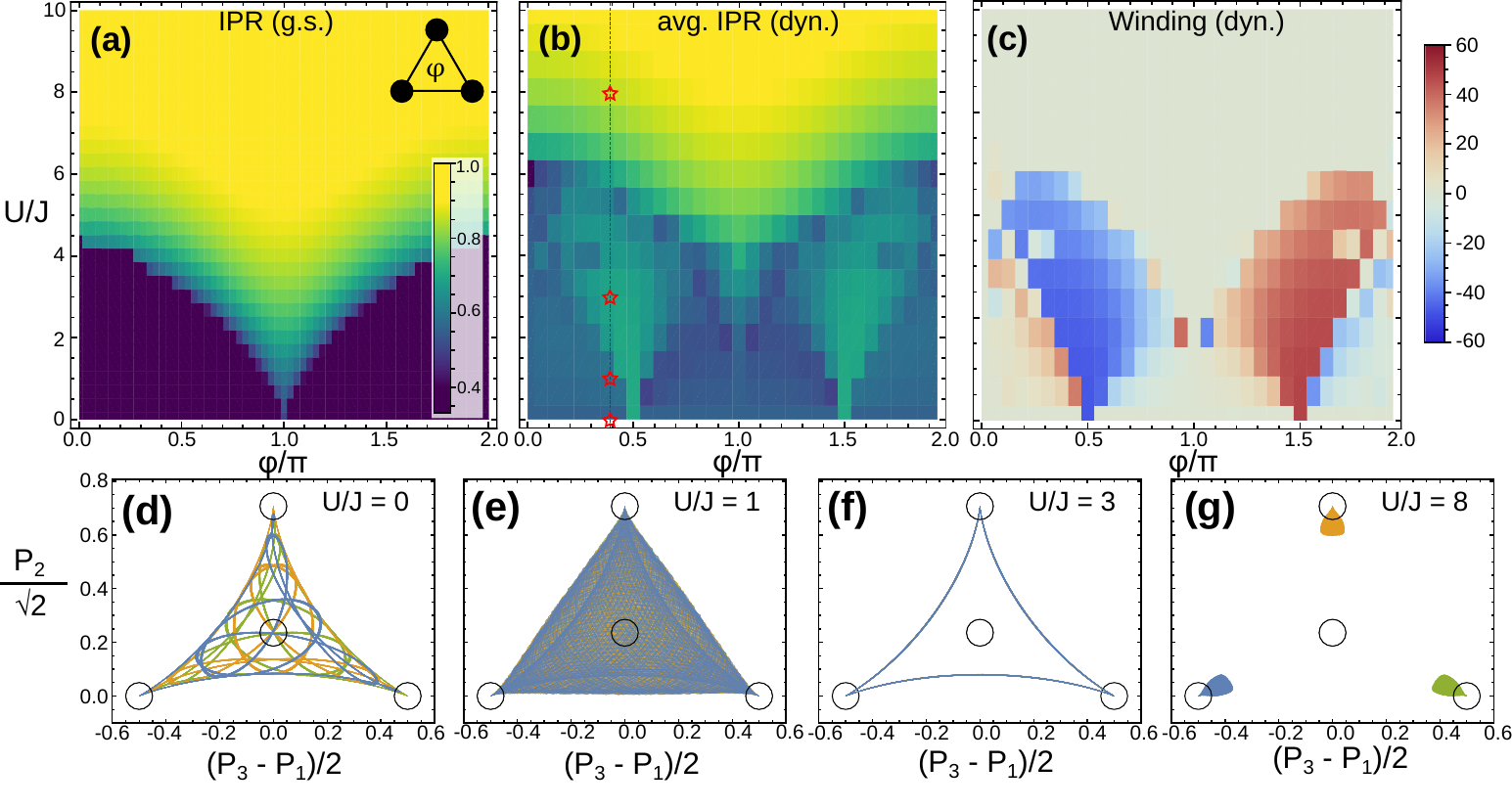}
	\centering
	\caption{\label{FIG:F1}
        \textbf{Numerically simulated equilibrium and dynamical properties of the 3-site Josephson ring.}
        \textbf{(a)}~Ground state inverse participation ratio (IPR) for the three-site system (depicted in the inset) with flux $\varphi$, inter-site hopping $J$, and on-site interaction $U$. Two regions appear: a self-trapped region at large $U/J$ and a region of delocalization at low $U/J$, with a $\varphi$-dependent self-trapping transition boundary. 
        \textbf{(b)}~Averaged dynamical IPR for the same system initialized with population at only one site. 
        \textbf{(c)}~Average winding of the center of mass about the 3-site loop under the conditions as in (b).
        In the non-equilibrium case (b,c), one finds two fan-shaped regions that are not present in the equilibrium scenario (a).
        These regions, which originate at $\varphi = \pi/2$ and $3\pi/2$ and grow (become stabilized) with increasing $U$, host modes that are localized (have increased IPR) and wind around the ring.
        \textbf{(d-g)}~Center-of-mass trajectory for excitations starting at each of the 3 initial sites (blue, green, orange) for $\varphi \approx 0.39\pi$ and increasing $U/J$ [dotted line and stars in \textbf{(b)}]. The trajectories are shown in the projected phase space of the site populations and population differences, where $P_j$ is the normalized population residing at site (oscillator) $j$.
        For $U/J=0$ (d), there is no net chiral motion, with the excitation following a closed trajectory. As $U/J$ increases, the trajectory first becomes chaotic (e), then enters a chiral, self-localized regime (f), and becomes fully self-trapped and immobile for large nonlinearities (g).
        }
\end{figure*}

Our experimental system consists of three feedback-coupled mechanical oscillators (i.e., an active mechanical lattice~\cite{Brandenbourger2019,Ilan-prop,Top-Morph-active,Veenstra2024}), described in Refs.~\cite{Anandwade-synthetic,Martello-Coexistence,Singhal-NH-AB,tian2023observation}. 
The oscillators serve as effective ``lattice sites,'' and they are physically uncoupled such that they do not naturally exchange energy (phonons).
We build a ``mechanical ring lattice'' by engineering artificial spring forces, based on real-time measurements and feedback~\cite{SuppMats}, that allow energy (phonons) to ``hop'' between the oscillators. 
Free from the restrictions of physical springs, we incorporate momentum-dependent forces to engineer complex hopping and an artificial gauge field. We additionally use feedback to introduce artificial nonlinearities~\cite{SuppMats}.

Our experiments operate deep in the classical regime, typically with more than Avogadro's number of phonons per oscillator. Still, they allow us to investigate phenomena stemming from the interplay of gauge fields and classical nonlinearities. The system of three discrete sites coupled in a ring geometry with an enclosed flux $\varphi$ is depicted in the Fig.~\pref{FIG:F1}{a} inset.
The dynamics of this system can be described by the classical nonlinear wave equation
\begin{align}
i \dot{\psi}_j = (\omega_0 - U |\psi_j|^2)\psi_j -J \sum_{k\neq j} e^{i \varphi_{kj}}\psi_k \ ,
\label{eq:Ham}
\end{align}
where the $\psi_j$ represent normalized wave amplitudes at the three oscillators, $\omega_0$ is the common bare angular frequency of the oscillators, $J/2\pi$ is the effective phonon hopping rate, and $\varphi_{23} = - \varphi_{32} \equiv \varphi$ (with $\varphi_{12} = \varphi_{31} = 0$) sets the flux in the loop. Finally, $U$ represents the tunable strength of the nonlinear self interactions (i.e., with $\sum_j |\psi_j|^2 = 1$, $U$ is the angular frequency shift experienced by an oscillator when it stores all of the energy).

Figure~\ref{FIG:F1} depicts the phases and phenomena that arise from Eq.~\ref{eq:Ham}, both in and out of equilibrium.
Figure~\pref{FIG:F1}{a} shows the
inverse participation ratio (IPR, $\sum_j |\Psi_j|^{4}$) of the ground state, which reveals a transition between perfectly delocalized states at low $U/J$ (dark blue, IPR of $1/3$) and regions of macroscopic self-trapping at large $U/J$ (yellow regions, IPR~$\sim$~1). The $\varphi$-dependence of the self-trapping phase boundary reflects the flux-tunability of the energy gap above the $U=0$ ground state~\cite{Trimer2_Cao_2015,Trimer-selftrap-regimes}.
We note that the delocalized ground states also display chiral currents~\cite{Roushan2017} for $\varphi \neq 0, \pi$, while the self-trapped regime supports no currents, chiral or otherwise.

The rest of Fig.~\ref{FIG:F1} [panels (b-g)] details the rich out-of-equilibrium behavior of the Josephson trimer, considering the case where population is initialized at a single site.
Figure~\pref{FIG:F1}{b} shows time-dependent IPR averaged over a long period ($\sim$84~$T_J$, with $T_J = 2\pi / J$) of dynamical evolution. Roughly speaking, three regions can be identified based on the time-averaged IPR. There is a self-trapped regime at large $U/J$, with a flux-dependent transition boundary that occurs at slightly larger $U/J$ values than in equilibrium. Below this, there are broad regions of low average IPR consistent with nearly random dynamics (the ``random'' state IPR $\approx 0.541$) within the trimer. Finally, there are two fan-like regions of increased IPR that originate as singular points at $\varphi = \pi/2$ and $3\pi/2$ and grow in their $\varphi$ extent as $U/J$ is increased, indicative of interaction-stabilized self-localized excitations. Figure~\pref{FIG:F1}{c} shows, for the same conditions as in (b), the net number of times that the centre-of-mass winds around the ring. As expected, there is no net winding in the regions associated with self-trapping and ``random'' dynamics.
But, in the fan-like regions, we find that the apparently self-localized excitations have significant winding in the clockwise (CW) and counter-clockwise (CCW) directions for $\varphi \sim \pi/2$ and $3\pi/2$, respectively.

Figure~\pref{FIG:F1}{d-g} show directly how an increasing $U/J$ influences the projected phase space dynamics.
For the chosen $\varphi$ value of $\approx 0.39 \pi$ [markers in (b)] and starting at individual sites, the dynamics with $U=0$ (d) relates to Lissajous curves. The addition of a small but non-zero $U/J$ (e) leads to dynamics uniformly filling the projected phase space, suggestive of chaotic dynamics (while the Josephson dimer is integrable, the undriven trimer supports chaos~\cite{Trimer1-selftrapping-etc}).
For moderate $U/J$ (f), the dynamics follows a singular path with perfect CCW winding.
Based on its high time-averaged IPR and chiral motion, we refer to this mode as a chiral solitary wave.
Finally, the phase space dynamics at large $U/J$ (g) remain effectively localized to the initial sites.

\begin{figure}[t!]
	\includegraphics[width=0.98\columnwidth]{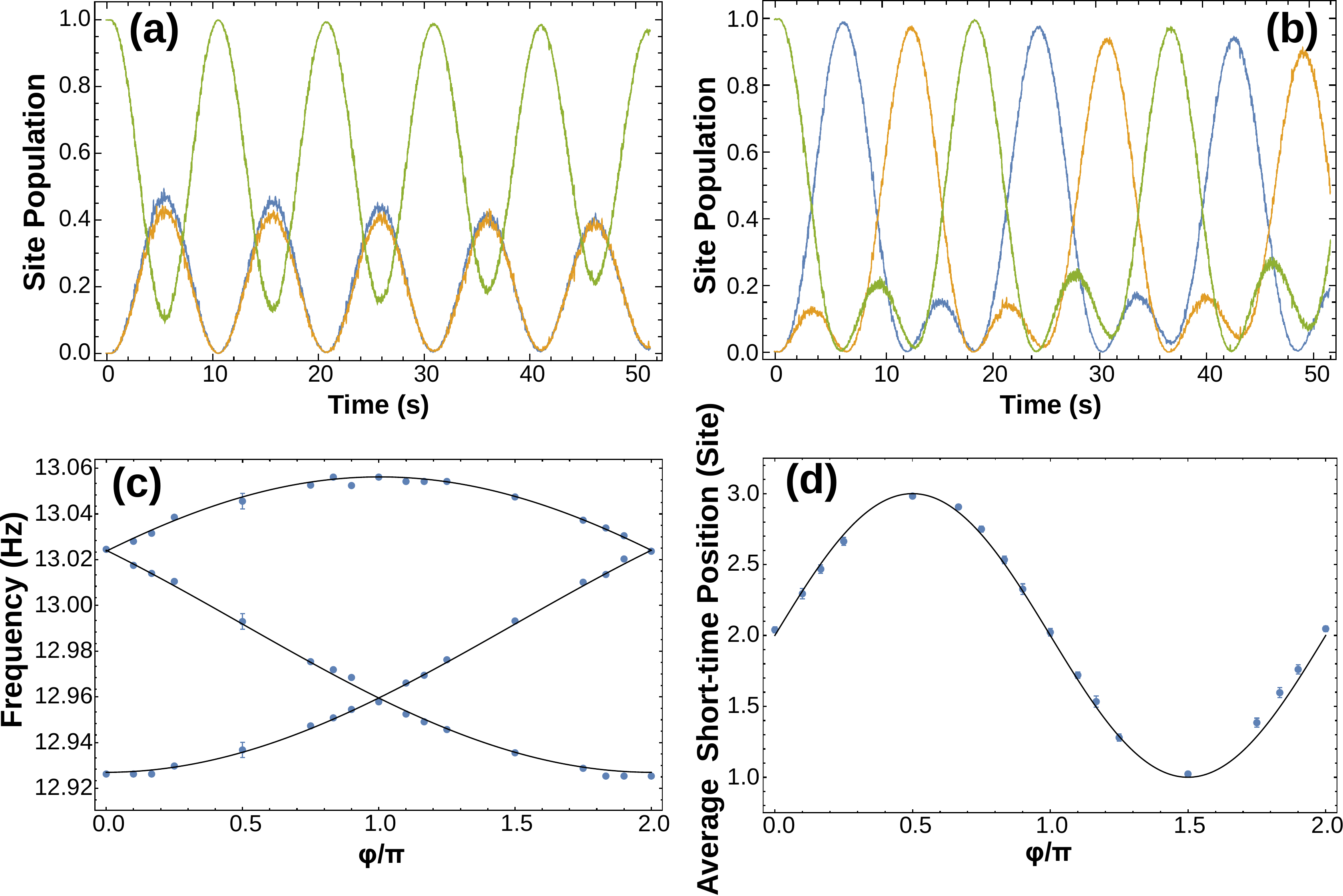}
	\centering
	\caption{\label{FIG:F2a}
		\textbf{Flux-driven chiral dynamics and spectral flow in a mechanical ring.}
        \textbf{(a,b)}~Normalized experimental dynamics of the mechanical energy, following initialization at oscillator 2 (green), for flux values of 0 and  $\pi/2$, respectively.
        \textbf{(c)}~Measured frequency spectrum vs. the applied loop flux $\varphi$. The normal mode frequencies are obtained by finding the peaks of a discrete Fourier transform based on 300~s of dynamics of the position variable for oscillator 2 ($X_2$).
        The solid black curve is the theoretical spectrum (for uniform $J/2\pi = 32.3(1)$~mHz and uniform bare frequency $\omega_0 = 12.9916(2)$~Hz).
        The representative experimental error bars (shown for $\varphi = \pi/2$) relate to the Fourier-limited experimental frequency uncertainty.
        \textbf{(d)}~Average position of the excitation at short evolution time (averaged over the period from $6.0 \pm 0.1$~s), serving as a measure of the wave packet velocity.
        For zero flux, under TRS, there is no net velocity. Maximally positive (clockwise) and negative (counterclockwise) displacements of the excitation energy are observed at flux values of $\pi/2$ and $3\pi/2$, respectively. The roughly sinusoidal dependence of the measured displacement on the applied flux $\varphi$ is in excellent agreement with the solid line theory curve, which is a no-free-parameter numerical simulation curve, using the values of $J$ as determined from panel \textbf{(c)}. Data error bars relate to the standard error of the mean of the displacement over the specified 300~s time window.
	}
\end{figure}

We begin our experiments in the linear regime ($U = 0)$, verifying the control of a tunable flux $\varphi$. The natural dynamics of phonons in a static system of spring-coupled masses - with only position-dependent forces - would be described by real-valued hoppings. This is similar to the predicament of other charge-neutral systems~\cite{AIDELSBURGER2018394} (e.g., photons~\cite{top-phot}
and neutral atoms~\cite{TopBands-Spielman}). Previously, the engineering of complex phonon hopping has been proposed by Floquet driving~\cite{salerno2016}, and it has been demonstrated with gyroscopic oscillators~\cite{GyroscopeMetamaterials}, by mathematical mappings in systems with enlarged unit cells~\cite{HuberPendulum,susstrunk2016classification,Huber2016}, and through dynamical optical modulation~\cite{Phonon-gaugefield}.
Here, we directly implement tunable complex hopping from $k \rightarrow j$
with phase $\varphi_{kj}$
by applying inter-oscillator forces of the form $F_j \propto (X_k \cos (\varphi_{kj}) + P_k \sin (\varphi_{kj}))$ [and $F_k \propto (X_j \cos (\varphi_{kj}) - P_j \sin (\varphi_{kj}))$ for the conjugate term], where $X_{k(j)}$ and $P_{k(j)}$ are the co-normalized positions and momenta of oscillators $k$ and $j$~\cite{SuppMats}.

Figure~\pref{FIG:F2a}{a,b} show the measured short-time dynamics of the normalized energy ($E_k \propto X_k^2 + P_k^2$) in our three-oscillator ring for flux values of $\varphi = 0$ and $\pi/2$. Initializing the energy at oscillator 2, we find that it (a) spreads symmetrically when no flux is applied and (b) runs around the ring in a chiral fashion for $\varphi = \pi/2$. Figure~\pref{FIG:F2a}{c} shows the experimental normal mode spectrum of the trimer as a function of $\varphi$, based on the peaks of the discrete Fourier transform of the $X_2$ dynamics over 300~s of evolution.
The measured spectrum winds over $2\pi$ and is consistent with a tunneling rate of $J/2\pi = 32.3(1)$~mHz and a common oscillator frequency of $f_0 = \omega_0/2\pi = 12.9916(2)$~Hz (black theory curve).

Figure~\pref{FIG:F2a}{d} shows how the chirality of the dynamics, observed by the short-time displacement from the initial site (site 2), depends on $\varphi$. We plot $\braket{k} = \frac{\sum_k k E_k}{\sum_k E_k}$ averaged over a small time window for a short evolution time ($6.0 \pm 0.1$~s).
The measured difference from $\braket{k}_0 = 2$, which is approximately proportional to the mean initial wave packet velocity, winds with the flux over $2\pi$. It takes non-zero values when TRS is broken ($\varphi \neq 0$ or $\pi$), with maximal flow in the CW direction when $\varphi \sim \pi/2$ and in the CCW direction when $\varphi \sim 3\pi/2$. The solid line is the no-free-parameter numerical simulation, based on the fit parameters extracted from (c).

\begin{figure*}[t!]	\includegraphics[width=1.85\columnwidth]
 {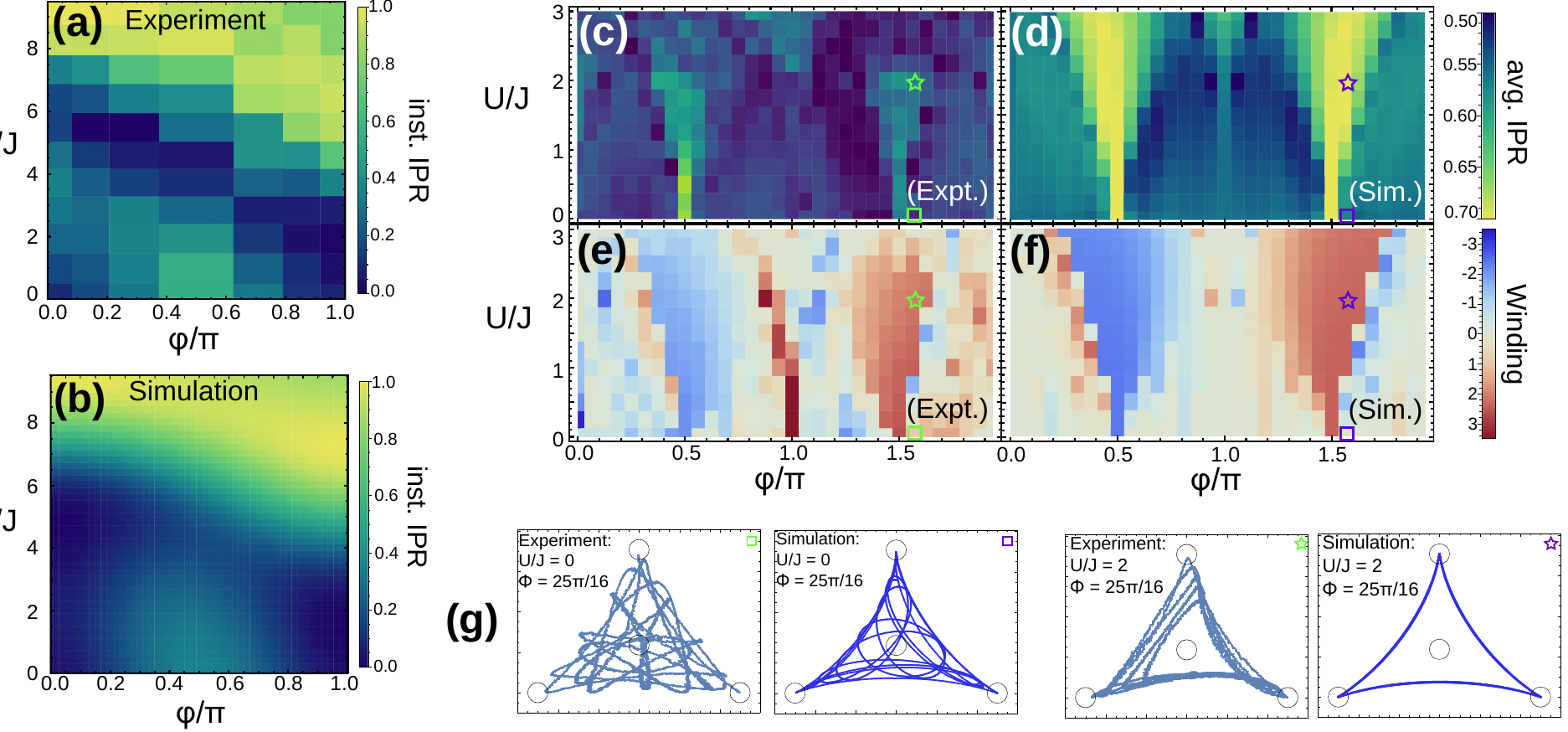}
	\centering
	\caption{\label{FIG:F3}
		\textbf{Observations of flux-dependent self-trapping and self-localized chiral excitations.}
        \textbf{(a,b)}~Instantaneous IPR after 1 tunneling time (16~s) for a tunneling rate of 8~mHz. As the nonlinear interaction increases, there is a transition from a dispersive regime to a fully self-localized regime of the system. Near $\pi/2$ flux, there appears to exist a region of self-localized behavior even for small $U/J$ values. This region in parameter space is examined further in (c-f).
        \textbf{(c,d)}~Average IPR for 120~s of time evolution. Numerics evaluated at experimental tunneling rate, self-interaction strength, and evolution time. Initially, the system is in a dispersive regime, with the excitation remaining localized only at $\pi/2$ and $3\pi/2$. As the nonlinear interaction strength increases, the localized region expands similarly to the winding.
        \textbf{(e,f)}~Average winding around the plaquette for 120~s of time evolution and $\sim$28~mHz tunneling rate. Numerics evaluated at experimental tunneling rate, self-interaction strength, and evolution time. At no non-linearity, the system only winds at $\pi/2$ and $3\pi/2$ flux. As the non-linear interaction increases, the region where the dynamics winds expands. The lines at $0$ and $\pi$ in the experimental data is a consequence of the winding being ill-defined at those points.
        \textbf{(g)}~Experimental and simulated trajectories of the center of mass around the trimer. Each circle on the edge represents the excitation being fully at one site, while the center circle represents all three sites being equally excited. A trajectory with no winding initially (square markers) becomes chiral with the addition of the nonlinear interaction (star markers).
        } 
\end{figure*}

We now explore how the dynamics is enriched by the addition of a tunable self-nonlinearity at each oscillator. 
We apply self-feedback forces of the form $F_k \propto (X_k^2 + P_k^2) X_k$~\cite{Anandwade-synthetic} at each oscillator, which give rise to the effective Hartree-like density-dependent potential term (or effective Kerr nonlinear term) $\propto |\psi_j|^2$ in Eq.~\ref{eq:Ham}.
Figure~\ref{FIG:F3} explores how an increasing $U/J$ affects the trimer's dynamical response.
In Fig.~\pref{FIG:F3}{a,b}, we first scan $U/J$ over a large range.
To allow us to explore the large interaction limit ($U \gg J$) while still adhering to technical limits on our feedback forcing, here we operate with a hopping rate $J$ that is four times smaller than in the case of Fig.~\ref{FIG:F2a} (i.e., here $J/2\pi \approx 8$~mHz). 
We again initially excite oscillator 2 and allow the system to evolve for some time. We measure the energy distribution over a time-window of $16.1 \pm 0.1$~s,
relating to the time when the population maximally spreads out for $U=0$.

Figure~\ref{FIG:F3}~(a,b) shows the experimentally measured and numerically predicted instantaneous IPR of the system after $16.1$~s as a function of $\varphi$ and the interaction-to-hopping ratio $U/J$. In general, we find that for all $\varphi$ values there is a transition from the non-interacting limit, where energy spreads out across the trimer, to a self-trapped regime where the energy remains stuck at oscillator 2. We observe the expected $\varphi$ dependence of the self-trapping transition, expected both in equilibrium and in the non-equilibrium response (see Fig.~\ref{FIG:F1}).
That is, unlike in the canonical two-mode case~\cite{Raghavan,Albiez-DirectJosephson,Zibold,Anandwade-synthetic}, the critical $U/J$ ratio required for self-trapping in the ring trimer is reduced due to frustration~\cite{Trimer2_Cao_2015,Trimer-selftrap-regimes,Jia-frustr}.
We also see signatures of a reduced spreading at small $U/J$ for $\varphi \sim \pi/2$, which we investigate in more detail.

Figure~\pref{FIG:F3}{c-f} shows the experimentally measured (c,e) and numerically predicted (d,f) behavior of the system across a smaller range of $U/J$ and with the response averaged over $120$~s of dynamics. To note, here we also operate with a larger value of $J/2\pi \approx 28$~mHz.
Looking at the (c) experimental time-averaged IPR, we see clear evidence for two regions of reduced IPR that originate at $\varphi = \pi/2$ and $3\pi/2$ and grow or fan out with increasing $U/J$. This is qualitatively consistent with the simulated response in (d) as well as Fig.~\ref{FIG:F1}. The emergent regions of interaction-stabilized localization seen in experiment are less robust than those in the simulations (i.e., they die off for $U/J \gtrsim 2$), likely due to experimental imperfections (e.g., noise or disorder in the parameter values, including loss and gain). Figure~\pref{FIG:F3}{e,f} examines the chirality of the measured and simulated dynamics through the net winding (number of center-of-mass windings around the ring). We find good agreement between experiment (e) and theory (f), showing evidence that the self-localized excitations are chiral, winding in the CW direction for $\varphi \sim \pi/2$ and in the CCW direction for $\varphi \sim 3\pi/2$.

Overall, these measurements indicate the observation of interaction-stabilized chiral excitations that are self-localized, consistent with the prediction from Fig.~\ref{FIG:F1} that the Aharonov-Bohm trimer with broken time-reversal symmetry supports emergent chiral solitary waves. In Fig.~\pref{FIG:F3}{g}, we additionally plot individual projected phase space trajectories (both experimental and simulations, for $\varphi = 25 \pi/16$) that illustrate directly how nonlinearities serve to stabilize chiral flow in the trimer.

Recalling one of the recent motivations for exploring nonlinear dynamics in an Aharonov-Bohm ring~\cite{Sacha-excited,Ohberg-gaugetheories,sacha-comment}, we note that the robust recurrent dynamics in our discrete 3-site lattice display some properties reminiscent of the predictions of Ref.~\cite{Sacha-excited}, which describes the emergence of excited state continuous time crystals in interacting flux rings. 
Looking forward, the active mechanics platform admits numerous ingredients that may be naturally added to this system, suggesting that excited state and even equilibrium classical time crystals~\cite{Sacha_2018,Oded-TC,yao-np2020-timecrystals,timecrystal-RMP,Hannaford_2022,Vitelli-Ising} may be explored in such a platform with the addition of noise, dissipation, non-reciprocity~\cite{time-crystal-nonrecip,Time-crystal-nonr}, long-ranged multi-body interactions~\cite{Kozin-LR}, dynamical gauge fields~\cite{Ohberg-gaugetheories}, and time-delayed interactions~\cite{Hemmerich-diss-TC}.
Beyond the pervasive physics of time crystals, the flexible architecture of active mechanics also presents new opportunities to explore myriad phenomena resulting from the interplay of nonlinearities and artificial gauge fields~\cite{Mukh-top-soliton,Mueller-bosons-ladder,Vekua-reversal,synth-flux-ladder,NonlinearModesGaugeFields}, as well as studies of density-dependent (dynamical) gauge fields~\cite{Edmonds-Gauge,Frolian2022,Faugno-dens} and lattice gauge theories in their classical limit~\cite{FredJlgt}.

\section{Acknowledgements}
This material (I.~V., A.~A., J.~Y., Y.~S., A.~G., and B.~G.) is based upon work supported by the National Science Foundation under grant No.~1945031. I.~V. and B.~G. acknowledge support from the AFOSR MURI program under agreement number FA9550-22-1-0339.
D.~L. acknowledges REU support from the National Science Foundation under grant No.~1950744.
J.~Y. and A.~G. acknowledge support by the John A. Gardner Undergraduate Research Award of the UIUC Department of Physics. J.~Y. additionally acknowledges support by the Simmons Undergraduate Summer Research Scholarship of the UIUC Department of Physics.
Y.~S. acknowledges support by the Jeremiah D. Sullivan Undergraduate Research Award of the UIUC Department of Physics.
T.~O. acknowledges support from JSPS KAKENHI Grant No. JP24K00548, JST PRESTO Grant No. JPMJPR2353, JST CREST Grant No. JPMJCR19T1.
E.~M. and H.~M.~P. are supported by the Royal Society via grants UF160112, URF\textbackslash R\textbackslash 221004, RGF\textbackslash EA\textbackslash 180121 and RGF\textbackslash 
 R1\textbackslash 180071, and by the Engineering and Physical Sciences Research Council (grant no. EP/W016141/1).
 E.~M. is supported also by PNRR MUR project PE0000023-NQSTI. We would like to thank Mike Gunn for useful discussions. 
This work was also supported by the BRIDGE Seed Fund for collaboration between the University of Birmingham and the University of Illinois at Urbana-Champaign.

\bibliographystyle{apsrev4-1}
\bibliography{3site}

\begin{thebibliography}{60}%
\makeatletter
\providecommand \@ifxundefined [1]{%
 \@ifx{#1\undefined}
}%
\providecommand \@ifnum [1]{%
 \ifnum #1\expandafter \@firstoftwo
 \else \expandafter \@secondoftwo
 \fi
}%
\providecommand \@ifx [1]{%
 \ifx #1\expandafter \@firstoftwo
 \else \expandafter \@secondoftwo
 \fi
}%
\providecommand \natexlab [1]{#1}%
\providecommand \enquote  [1]{``#1''}%
\providecommand \bibnamefont  [1]{#1}%
\providecommand \bibfnamefont [1]{#1}%
\providecommand \citenamefont [1]{#1}%
\providecommand \href@noop [0]{\@secondoftwo}%
\providecommand \href [0]{\begingroup \@sanitize@url \@href}%
\providecommand \@href[1]{\@@startlink{#1}\@@href}%
\providecommand \@@href[1]{\endgroup#1\@@endlink}%
\providecommand \@sanitize@url [0]{\catcode `\\12\catcode `\$12\catcode
  `\&12\catcode `\#12\catcode `\^12\catcode `\_12\catcode `\%12\relax}%
\providecommand \@@startlink[1]{}%
\providecommand \@@endlink[0]{}%
\providecommand \url  [0]{\begingroup\@sanitize@url \@url }%
\providecommand \@url [1]{\endgroup\@href {#1}{\urlprefix }}%
\providecommand \urlprefix  [0]{URL }%
\providecommand \Eprint [0]{\href }%
\providecommand \doibase [0]{http://dx.doi.org/}%
\providecommand \selectlanguage [0]{\@gobble}%
\providecommand \bibinfo  [0]{\@secondoftwo}%
\providecommand \bibfield  [0]{\@secondoftwo}%
\providecommand \translation [1]{[#1]}%
\providecommand \BibitemOpen [0]{}%
\providecommand \bibitemStop [0]{}%
\providecommand \bibitemNoStop [0]{.\EOS\space}%
\providecommand \EOS [0]{\spacefactor3000\relax}%
\providecommand \BibitemShut  [1]{\csname bibitem#1\endcsname}%
\let\auto@bib@innerbib\@empty
\bibitem [{\citenamefont {von Klitzing}(1986)}]{vonKlitzing-IQHE}%
  \BibitemOpen
  \bibfield  {author} {\bibinfo {author} {\bibfnamefont {K.}~\bibnamefont {von
  Klitzing}},\ }\href {\doibase 10.1103/RevModPhys.58.519} {\bibfield
  {journal} {\bibinfo  {journal} {Rev. Mod. Phys.}\ }\textbf {\bibinfo {volume}
  {58}},\ \bibinfo {pages} {519} (\bibinfo {year} {1986})}\BibitemShut
  {NoStop}%
\bibitem [{\citenamefont {Stormer}\ \emph {et~al.}(1999)\citenamefont
  {Stormer}, \citenamefont {Tsui},\ and\ \citenamefont
  {Gossard}}]{Stormer-FQHE}%
  \BibitemOpen
  \bibfield  {author} {\bibinfo {author} {\bibfnamefont {H.~L.}\ \bibnamefont
  {Stormer}}, \bibinfo {author} {\bibfnamefont {D.~C.}\ \bibnamefont {Tsui}}, \
  and\ \bibinfo {author} {\bibfnamefont {A.~C.}\ \bibnamefont {Gossard}},\
  }\href {\doibase 10.1103/RevModPhys.71.S298} {\bibfield  {journal} {\bibinfo
  {journal} {Rev. Mod. Phys.}\ }\textbf {\bibinfo {volume} {71}},\ \bibinfo
  {pages} {S298} (\bibinfo {year} {1999})}\BibitemShut {NoStop}%
\bibitem [{\citenamefont {Wei}\ and\ \citenamefont
  {Mueller}(2014)}]{Mueller-bosons-ladder}%
  \BibitemOpen
  \bibfield  {author} {\bibinfo {author} {\bibfnamefont {R.}~\bibnamefont
  {Wei}}\ and\ \bibinfo {author} {\bibfnamefont {E.~J.}\ \bibnamefont
  {Mueller}},\ }\href {\doibase 10.1103/PhysRevA.89.063617} {\bibfield
  {journal} {\bibinfo  {journal} {Phys. Rev. A}\ }\textbf {\bibinfo {volume}
  {89}},\ \bibinfo {pages} {063617} (\bibinfo {year} {2014})}\BibitemShut
  {NoStop}%
\bibitem [{\citenamefont {Li}\ \emph {et~al.}(2023)\citenamefont {Li},
  \citenamefont {Du}, \citenamefont {Wang}, \citenamefont {Liang},
  \citenamefont {Xiao}, \citenamefont {Yi}, \citenamefont {Ma},\ and\
  \citenamefont {Jia}}]{Jia-frustr}%
  \BibitemOpen
  \bibfield  {author} {\bibinfo {author} {\bibfnamefont {Y.}~\bibnamefont
  {Li}}, \bibinfo {author} {\bibfnamefont {H.}~\bibnamefont {Du}}, \bibinfo
  {author} {\bibfnamefont {Y.}~\bibnamefont {Wang}}, \bibinfo {author}
  {\bibfnamefont {J.}~\bibnamefont {Liang}}, \bibinfo {author} {\bibfnamefont
  {L.}~\bibnamefont {Xiao}}, \bibinfo {author} {\bibfnamefont {W.}~\bibnamefont
  {Yi}}, \bibinfo {author} {\bibfnamefont {J.}~\bibnamefont {Ma}}, \ and\
  \bibinfo {author} {\bibfnamefont {S.}~\bibnamefont {Jia}},\ }\href {\doibase
  10.1038/s41467-023-43204-3} {\bibfield  {journal} {\bibinfo  {journal}
  {Nature Communications}\ }\textbf {\bibinfo {volume} {14}},\ \bibinfo {pages}
  {7560} (\bibinfo {year} {2023})}\BibitemShut {NoStop}%
\bibitem [{\citenamefont {Aidelsburger}\ \emph {et~al.}(2018)\citenamefont
  {Aidelsburger}, \citenamefont {Nascimbene},\ and\ \citenamefont
  {Goldman}}]{AIDELSBURGER2018394}%
  \BibitemOpen
  \bibfield  {author} {\bibinfo {author} {\bibfnamefont {M.}~\bibnamefont
  {Aidelsburger}}, \bibinfo {author} {\bibfnamefont {S.}~\bibnamefont
  {Nascimbene}}, \ and\ \bibinfo {author} {\bibfnamefont {N.}~\bibnamefont
  {Goldman}},\ }\href {\doibase https://doi.org/10.1016/j.crhy.2018.03.002}
  {\bibfield  {journal} {\bibinfo  {journal} {Comptes Rendus Physique}\
  }\textbf {\bibinfo {volume} {19}},\ \bibinfo {pages} {394} (\bibinfo {year}
  {2018})}\BibitemShut {NoStop}%
\bibitem [{\citenamefont {{Zakharov}}\ and\ \citenamefont
  {{Shabat}}(1972)}]{soliton-1}%
  \BibitemOpen
  \bibfield  {author} {\bibinfo {author} {\bibfnamefont {V.~E.}\ \bibnamefont
  {{Zakharov}}}\ and\ \bibinfo {author} {\bibfnamefont {A.~B.}\ \bibnamefont
  {{Shabat}}},\ }\href@noop {} {\bibfield  {journal} {\bibinfo  {journal}
  {Soviet Journal of Experimental and Theoretical Physics}\ }\textbf {\bibinfo
  {volume} {34}},\ \bibinfo {pages} {62} (\bibinfo {year} {1972})}\BibitemShut
  {NoStop}%
\bibitem [{\citenamefont {Morgan}\ \emph {et~al.}(1997)\citenamefont {Morgan},
  \citenamefont {Ballagh},\ and\ \citenamefont {Burnett}}]{soliton-burnett}%
  \BibitemOpen
  \bibfield  {author} {\bibinfo {author} {\bibfnamefont {S.~A.}\ \bibnamefont
  {Morgan}}, \bibinfo {author} {\bibfnamefont {R.~J.}\ \bibnamefont {Ballagh}},
  \ and\ \bibinfo {author} {\bibfnamefont {K.}~\bibnamefont {Burnett}},\ }\href
  {\doibase 10.1103/PhysRevA.55.4338} {\bibfield  {journal} {\bibinfo
  {journal} {Phys. Rev. A}\ }\textbf {\bibinfo {volume} {55}},\ \bibinfo
  {pages} {4338} (\bibinfo {year} {1997})}\BibitemShut {NoStop}%
\bibitem [{\citenamefont {Mitchell}\ \emph {et~al.}(1996)\citenamefont
  {Mitchell}, \citenamefont {Chen}, \citenamefont {Shih},\ and\ \citenamefont
  {Segev}}]{Self-trap-1}%
  \BibitemOpen
  \bibfield  {author} {\bibinfo {author} {\bibfnamefont {M.}~\bibnamefont
  {Mitchell}}, \bibinfo {author} {\bibfnamefont {Z.}~\bibnamefont {Chen}},
  \bibinfo {author} {\bibfnamefont {M.-f.}\ \bibnamefont {Shih}}, \ and\
  \bibinfo {author} {\bibfnamefont {M.}~\bibnamefont {Segev}},\ }\href
  {\doibase 10.1103/PhysRevLett.77.490} {\bibfield  {journal} {\bibinfo
  {journal} {Phys. Rev. Lett.}\ }\textbf {\bibinfo {volume} {77}},\ \bibinfo
  {pages} {490} (\bibinfo {year} {1996})}\BibitemShut {NoStop}%
\bibitem [{\citenamefont {Fleischer}\ \emph {et~al.}(2003)\citenamefont
  {Fleischer}, \citenamefont {Carmon}, \citenamefont {Segev}, \citenamefont
  {Efremidis},\ and\ \citenamefont {Christodoulides}}]{Soliton-Segev-Christo}%
  \BibitemOpen
  \bibfield  {author} {\bibinfo {author} {\bibfnamefont {J.~W.}\ \bibnamefont
  {Fleischer}}, \bibinfo {author} {\bibfnamefont {T.}~\bibnamefont {Carmon}},
  \bibinfo {author} {\bibfnamefont {M.}~\bibnamefont {Segev}}, \bibinfo
  {author} {\bibfnamefont {N.~K.}\ \bibnamefont {Efremidis}}, \ and\ \bibinfo
  {author} {\bibfnamefont {D.~N.}\ \bibnamefont {Christodoulides}},\ }\href
  {\doibase 10.1103/PhysRevLett.90.023902} {\bibfield  {journal} {\bibinfo
  {journal} {Phys. Rev. Lett.}\ }\textbf {\bibinfo {volume} {90}},\ \bibinfo
  {pages} {023902} (\bibinfo {year} {2003})}\BibitemShut {NoStop}%
\bibitem [{\citenamefont {Raghavan}\ \emph {et~al.}(1999)\citenamefont
  {Raghavan}, \citenamefont {Smerzi}, \citenamefont {Fantoni},\ and\
  \citenamefont {Shenoy}}]{Raghavan}%
  \BibitemOpen
  \bibfield  {author} {\bibinfo {author} {\bibfnamefont {S.}~\bibnamefont
  {Raghavan}}, \bibinfo {author} {\bibfnamefont {A.}~\bibnamefont {Smerzi}},
  \bibinfo {author} {\bibfnamefont {S.}~\bibnamefont {Fantoni}}, \ and\
  \bibinfo {author} {\bibfnamefont {S.~R.}\ \bibnamefont {Shenoy}},\ }\href
  {\doibase 10.1103/PhysRevA.59.620} {\bibfield  {journal} {\bibinfo  {journal}
  {Phys. Rev. A}\ }\textbf {\bibinfo {volume} {59}},\ \bibinfo {pages} {620}
  (\bibinfo {year} {1999})}\BibitemShut {NoStop}%
\bibitem [{\citenamefont {Eilbeck}\ \emph {et~al.}(1985)\citenamefont
  {Eilbeck}, \citenamefont {Lomdahl},\ and\ \citenamefont
  {Scott}}]{Trimer-DST}%
  \BibitemOpen
  \bibfield  {author} {\bibinfo {author} {\bibfnamefont {J.}~\bibnamefont
  {Eilbeck}}, \bibinfo {author} {\bibfnamefont {P.}~\bibnamefont {Lomdahl}}, \
  and\ \bibinfo {author} {\bibfnamefont {A.}~\bibnamefont {Scott}},\ }\href
  {\doibase https://doi.org/10.1016/0167-2789(85)90012-0} {\bibfield  {journal}
  {\bibinfo  {journal} {Physica D: Nonlinear Phenomena}\ }\textbf {\bibinfo
  {volume} {16}},\ \bibinfo {pages} {318} (\bibinfo {year} {1985})}\BibitemShut
  {NoStop}%
\bibitem [{\citenamefont {Johansson}(2004)}]{Ham-Hopf-NEWMODE}%
  \BibitemOpen
  \bibfield  {author} {\bibinfo {author} {\bibfnamefont {M.}~\bibnamefont
  {Johansson}},\ }\href {\doibase 10.1088/0305-4470/37/6/017} {\bibfield
  {journal} {\bibinfo  {journal} {Journal of Physics A: Mathematical and
  General}\ }\textbf {\bibinfo {volume} {37}},\ \bibinfo {pages} {2201}
  (\bibinfo {year} {2004})}\BibitemShut {NoStop}%
\bibitem [{\citenamefont {Nemoto}\ \emph {et~al.}(2000)\citenamefont {Nemoto},
  \citenamefont {Holmes}, \citenamefont {Milburn},\ and\ \citenamefont
  {Munro}}]{Trimer0-munro}%
  \BibitemOpen
  \bibfield  {author} {\bibinfo {author} {\bibfnamefont {K.}~\bibnamefont
  {Nemoto}}, \bibinfo {author} {\bibfnamefont {C.~A.}\ \bibnamefont {Holmes}},
  \bibinfo {author} {\bibfnamefont {G.~J.}\ \bibnamefont {Milburn}}, \ and\
  \bibinfo {author} {\bibfnamefont {W.~J.}\ \bibnamefont {Munro}},\ }\href
  {\doibase 10.1103/PhysRevA.63.013604} {\bibfield  {journal} {\bibinfo
  {journal} {Phys. Rev. A}\ }\textbf {\bibinfo {volume} {63}},\ \bibinfo
  {pages} {013604} (\bibinfo {year} {2000})}\BibitemShut {NoStop}%
\bibitem [{\citenamefont {Franzosi}\ and\ \citenamefont
  {Penna}(2003)}]{Trimer1-selftrapping-etc}%
  \BibitemOpen
  \bibfield  {author} {\bibinfo {author} {\bibfnamefont {R.}~\bibnamefont
  {Franzosi}}\ and\ \bibinfo {author} {\bibfnamefont {V.}~\bibnamefont
  {Penna}},\ }\href {\doibase 10.1103/PhysRevE.67.046227} {\bibfield  {journal}
  {\bibinfo  {journal} {Phys. Rev. E}\ }\textbf {\bibinfo {volume} {67}},\
  \bibinfo {pages} {046227} (\bibinfo {year} {2003})}\BibitemShut {NoStop}%
\bibitem [{\citenamefont {Cao}\ \emph {et~al.}(2015)\citenamefont {Cao},
  \citenamefont {Wang},\ and\ \citenamefont {Fu}}]{Trimer2_Cao_2015}%
  \BibitemOpen
  \bibfield  {author} {\bibinfo {author} {\bibfnamefont {H.}~\bibnamefont
  {Cao}}, \bibinfo {author} {\bibfnamefont {Q.}~\bibnamefont {Wang}}, \ and\
  \bibinfo {author} {\bibfnamefont {L.-B.}\ \bibnamefont {Fu}},\ }\href
  {\doibase 10.1088/1054-660X/25/6/065501} {\bibfield  {journal} {\bibinfo
  {journal} {Laser Physics}\ }\textbf {\bibinfo {volume} {25}},\ \bibinfo
  {pages} {065501} (\bibinfo {year} {2015})}\BibitemShut {NoStop}%
\bibitem [{\citenamefont {Arwas}\ \emph {et~al.}(2014)\citenamefont {Arwas},
  \citenamefont {Vardi},\ and\ \citenamefont
  {Cohen}}]{Trimer-selftrap-regimes}%
  \BibitemOpen
  \bibfield  {author} {\bibinfo {author} {\bibfnamefont {G.}~\bibnamefont
  {Arwas}}, \bibinfo {author} {\bibfnamefont {A.}~\bibnamefont {Vardi}}, \ and\
  \bibinfo {author} {\bibfnamefont {D.}~\bibnamefont {Cohen}},\ }\href
  {\doibase 10.1103/PhysRevA.89.013601} {\bibfield  {journal} {\bibinfo
  {journal} {Phys. Rev. A}\ }\textbf {\bibinfo {volume} {89}},\ \bibinfo
  {pages} {013601} (\bibinfo {year} {2014})}\BibitemShut {NoStop}%
\bibitem [{\citenamefont {Gallemí}\ \emph {et~al.}(2015)\citenamefont
  {Gallemí}, \citenamefont {Guilleumas}, \citenamefont {Martorell},
  \citenamefont {Mayol}, \citenamefont {Polls},\ and\ \citenamefont
  {Juliá-Díaz}}]{3-site-BHH}%
  \BibitemOpen
  \bibfield  {author} {\bibinfo {author} {\bibfnamefont {A.}~\bibnamefont
  {Gallemí}}, \bibinfo {author} {\bibfnamefont {M.}~\bibnamefont
  {Guilleumas}}, \bibinfo {author} {\bibfnamefont {J.}~\bibnamefont
  {Martorell}}, \bibinfo {author} {\bibfnamefont {R.}~\bibnamefont {Mayol}},
  \bibinfo {author} {\bibfnamefont {A.}~\bibnamefont {Polls}}, \ and\ \bibinfo
  {author} {\bibfnamefont {B.}~\bibnamefont {Juliá-Díaz}},\ }\href {\doibase
  10.1088/1367-2630/17/7/073014} {\bibfield  {journal} {\bibinfo  {journal}
  {New Journal of Physics}\ }\textbf {\bibinfo {volume} {17}},\ \bibinfo
  {pages} {073014} (\bibinfo {year} {2015})}\BibitemShut {NoStop}%
\bibitem [{\citenamefont {Pizzi}\ \emph {et~al.}(2019)\citenamefont {Pizzi},
  \citenamefont {Dolcini},\ and\ \citenamefont {Le~Hur}}]{BHH-LeHur-synch}%
  \BibitemOpen
  \bibfield  {author} {\bibinfo {author} {\bibfnamefont {A.}~\bibnamefont
  {Pizzi}}, \bibinfo {author} {\bibfnamefont {F.}~\bibnamefont {Dolcini}}, \
  and\ \bibinfo {author} {\bibfnamefont {K.}~\bibnamefont {Le~Hur}},\ }\href
  {\doibase 10.1103/PhysRevB.99.094301} {\bibfield  {journal} {\bibinfo
  {journal} {Phys. Rev. B}\ }\textbf {\bibinfo {volume} {99}},\ \bibinfo
  {pages} {094301} (\bibinfo {year} {2019})}\BibitemShut {NoStop}%
\bibitem [{\citenamefont {Mukherjee}\ and\ \citenamefont
  {Rechtsman}(2020)}]{Mukh-top-soliton}%
  \BibitemOpen
  \bibfield  {author} {\bibinfo {author} {\bibfnamefont {S.}~\bibnamefont
  {Mukherjee}}\ and\ \bibinfo {author} {\bibfnamefont {M.~C.}\ \bibnamefont
  {Rechtsman}},\ }\href {\doibase 10.1126/science.aba8725} {\bibfield
  {journal} {\bibinfo  {journal} {Science}\ }\textbf {\bibinfo {volume}
  {368}},\ \bibinfo {pages} {856} (\bibinfo {year} {2020})}\BibitemShut
  {NoStop}%
\bibitem [{\citenamefont {Kartashov}\ and\ \citenamefont
  {Konotop}(2020)}]{NonlinearModesGaugeFields}%
  \BibitemOpen
  \bibfield  {author} {\bibinfo {author} {\bibfnamefont {Y.~V.}\ \bibnamefont
  {Kartashov}}\ and\ \bibinfo {author} {\bibfnamefont {V.~V.}\ \bibnamefont
  {Konotop}},\ }\href {\doibase 10.1103/PhysRevLett.125.054101} {\bibfield
  {journal} {\bibinfo  {journal} {Phys. Rev. Lett.}\ }\textbf {\bibinfo
  {volume} {125}},\ \bibinfo {pages} {054101} (\bibinfo {year}
  {2020})}\BibitemShut {NoStop}%
\bibitem [{\citenamefont {Wilczek}(2012)}]{Wilczek-TC}%
  \BibitemOpen
  \bibfield  {author} {\bibinfo {author} {\bibfnamefont {F.}~\bibnamefont
  {Wilczek}},\ }\href {\doibase 10.1103/PhysRevLett.109.160401} {\bibfield
  {journal} {\bibinfo  {journal} {Phys. Rev. Lett.}\ }\textbf {\bibinfo
  {volume} {109}},\ \bibinfo {pages} {160401} (\bibinfo {year}
  {2012})}\BibitemShut {NoStop}%
\bibitem [{\citenamefont {Syrwid}\ \emph {et~al.}(2017)\citenamefont {Syrwid},
  \citenamefont {Zakrzewski},\ and\ \citenamefont {Sacha}}]{Sacha-excited}%
  \BibitemOpen
  \bibfield  {author} {\bibinfo {author} {\bibfnamefont {A.}~\bibnamefont
  {Syrwid}}, \bibinfo {author} {\bibfnamefont {J.}~\bibnamefont {Zakrzewski}},
  \ and\ \bibinfo {author} {\bibfnamefont {K.}~\bibnamefont {Sacha}},\ }\href
  {\doibase 10.1103/PhysRevLett.119.250602} {\bibfield  {journal} {\bibinfo
  {journal} {Phys. Rev. Lett.}\ }\textbf {\bibinfo {volume} {119}},\ \bibinfo
  {pages} {250602} (\bibinfo {year} {2017})}\BibitemShut {NoStop}%
\bibitem [{\citenamefont {\"Ohberg}\ and\ \citenamefont
  {Wright}(2019)}]{Ohberg-gaugetheories}%
  \BibitemOpen
  \bibfield  {author} {\bibinfo {author} {\bibfnamefont {P.}~\bibnamefont
  {\"Ohberg}}\ and\ \bibinfo {author} {\bibfnamefont {E.~M.}\ \bibnamefont
  {Wright}},\ }\href {\doibase 10.1103/PhysRevLett.123.250402} {\bibfield
  {journal} {\bibinfo  {journal} {Phys. Rev. Lett.}\ }\textbf {\bibinfo
  {volume} {123}},\ \bibinfo {pages} {250402} (\bibinfo {year}
  {2019})}\BibitemShut {NoStop}%
\bibitem [{\citenamefont {Syrwid}\ \emph {et~al.}(2020)\citenamefont {Syrwid},
  \citenamefont {Kosior},\ and\ \citenamefont {Sacha}}]{sacha-comment}%
  \BibitemOpen
  \bibfield  {author} {\bibinfo {author} {\bibfnamefont {A.}~\bibnamefont
  {Syrwid}}, \bibinfo {author} {\bibfnamefont {A.}~\bibnamefont {Kosior}}, \
  and\ \bibinfo {author} {\bibfnamefont {K.}~\bibnamefont {Sacha}},\ }\href
  {\doibase 10.1103/PhysRevLett.124.178901} {\bibfield  {journal} {\bibinfo
  {journal} {Phys. Rev. Lett.}\ }\textbf {\bibinfo {volume} {124}},\ \bibinfo
  {pages} {178901} (\bibinfo {year} {2020})}\BibitemShut {NoStop}%
\bibitem [{\citenamefont {Brandenbourger}\ \emph {et~al.}(2019)\citenamefont
  {Brandenbourger}, \citenamefont {Locsin}, \citenamefont {Lerner},\ and\
  \citenamefont {Coulais}}]{Brandenbourger2019}%
  \BibitemOpen
  \bibfield  {author} {\bibinfo {author} {\bibfnamefont {M.}~\bibnamefont
  {Brandenbourger}}, \bibinfo {author} {\bibfnamefont {X.}~\bibnamefont
  {Locsin}}, \bibinfo {author} {\bibfnamefont {E.}~\bibnamefont {Lerner}}, \
  and\ \bibinfo {author} {\bibfnamefont {C.}~\bibnamefont {Coulais}},\ }\href
  {\doibase 10.1038/s41467-019-12599-3} {\bibfield  {journal} {\bibinfo
  {journal} {Nature Communications}\ }\textbf {\bibinfo {volume} {10}},\
  \bibinfo {pages} {4608} (\bibinfo {year} {2019})}\BibitemShut {NoStop}%
\bibitem [{\citenamefont {Sirota}\ \emph {et~al.}(2020)\citenamefont {Sirota},
  \citenamefont {Ilan}, \citenamefont {Shokef},\ and\ \citenamefont
  {Lahini}}]{Ilan-prop}%
  \BibitemOpen
  \bibfield  {author} {\bibinfo {author} {\bibfnamefont {L.}~\bibnamefont
  {Sirota}}, \bibinfo {author} {\bibfnamefont {R.}~\bibnamefont {Ilan}},
  \bibinfo {author} {\bibfnamefont {Y.}~\bibnamefont {Shokef}}, \ and\ \bibinfo
  {author} {\bibfnamefont {Y.}~\bibnamefont {Lahini}},\ }\href {\doibase
  10.1103/PhysRevLett.125.256802} {\bibfield  {journal} {\bibinfo  {journal}
  {Phys. Rev. Lett.}\ }\textbf {\bibinfo {volume} {125}},\ \bibinfo {pages}
  {256802} (\bibinfo {year} {2020})}\BibitemShut {NoStop}%
\bibitem [{\citenamefont {Wang}\ \emph {et~al.}(2022)\citenamefont {Wang},
  \citenamefont {Wang},\ and\ \citenamefont {Ma}}]{Top-Morph-active}%
  \BibitemOpen
  \bibfield  {author} {\bibinfo {author} {\bibfnamefont {W.}~\bibnamefont
  {Wang}}, \bibinfo {author} {\bibfnamefont {X.}~\bibnamefont {Wang}}, \ and\
  \bibinfo {author} {\bibfnamefont {G.}~\bibnamefont {Ma}},\ }\href {\doibase
  10.1038/s41586-022-04929-1} {\bibfield  {journal} {\bibinfo  {journal}
  {Nature}\ }\textbf {\bibinfo {volume} {608}},\ \bibinfo {pages} {50}
  (\bibinfo {year} {2022})}\BibitemShut {NoStop}%
\bibitem [{\citenamefont {Veenstra}\ \emph {et~al.}(2024)\citenamefont
  {Veenstra}, \citenamefont {Gamayun}, \citenamefont {Guo}, \citenamefont
  {Sarvi}, \citenamefont {Meinersen},\ and\ \citenamefont
  {Coulais}}]{Veenstra2024}%
  \BibitemOpen
  \bibfield  {author} {\bibinfo {author} {\bibfnamefont {J.}~\bibnamefont
  {Veenstra}}, \bibinfo {author} {\bibfnamefont {O.}~\bibnamefont {Gamayun}},
  \bibinfo {author} {\bibfnamefont {X.}~\bibnamefont {Guo}}, \bibinfo {author}
  {\bibfnamefont {A.}~\bibnamefont {Sarvi}}, \bibinfo {author} {\bibfnamefont
  {C.~V.}\ \bibnamefont {Meinersen}}, \ and\ \bibinfo {author} {\bibfnamefont
  {C.}~\bibnamefont {Coulais}},\ }\href {\doibase 10.1038/s41586-024-07097-6}
  {\bibfield  {journal} {\bibinfo  {journal} {Nature}\ }\textbf {\bibinfo
  {volume} {627}},\ \bibinfo {pages} {528} (\bibinfo {year}
  {2024})}\BibitemShut {NoStop}%
\bibitem [{\citenamefont {Anandwade}\ \emph {et~al.}(2023)\citenamefont
  {Anandwade}, \citenamefont {Singhal}, \citenamefont {Paladugu}, \citenamefont
  {Martello}, \citenamefont {Castle}, \citenamefont {Agrawal}, \citenamefont
  {Carlson}, \citenamefont {Battle-McDonald}, \citenamefont {Ozawa},
  \citenamefont {Price},\ and\ \citenamefont {Gadway}}]{Anandwade-synthetic}%
  \BibitemOpen
  \bibfield  {author} {\bibinfo {author} {\bibfnamefont {R.}~\bibnamefont
  {Anandwade}}, \bibinfo {author} {\bibfnamefont {Y.}~\bibnamefont {Singhal}},
  \bibinfo {author} {\bibfnamefont {S.~N.~M.}\ \bibnamefont {Paladugu}},
  \bibinfo {author} {\bibfnamefont {E.}~\bibnamefont {Martello}}, \bibinfo
  {author} {\bibfnamefont {M.}~\bibnamefont {Castle}}, \bibinfo {author}
  {\bibfnamefont {S.}~\bibnamefont {Agrawal}}, \bibinfo {author} {\bibfnamefont
  {E.}~\bibnamefont {Carlson}}, \bibinfo {author} {\bibfnamefont
  {C.}~\bibnamefont {Battle-McDonald}}, \bibinfo {author} {\bibfnamefont
  {T.}~\bibnamefont {Ozawa}}, \bibinfo {author} {\bibfnamefont {H.~M.}\
  \bibnamefont {Price}}, \ and\ \bibinfo {author} {\bibfnamefont
  {B.}~\bibnamefont {Gadway}},\ }\href {\doibase 10.1103/PhysRevA.108.012221}
  {\bibfield  {journal} {\bibinfo  {journal} {Phys. Rev. A}\ }\textbf {\bibinfo
  {volume} {108}},\ \bibinfo {pages} {012221} (\bibinfo {year}
  {2023})}\BibitemShut {NoStop}%
\bibitem [{\citenamefont {Martello}\ \emph {et~al.}(2023)\citenamefont
  {Martello}, \citenamefont {Singhal}, \citenamefont {Gadway}, \citenamefont
  {Ozawa},\ and\ \citenamefont {Price}}]{Martello-Coexistence}%
  \BibitemOpen
  \bibfield  {author} {\bibinfo {author} {\bibfnamefont {E.}~\bibnamefont
  {Martello}}, \bibinfo {author} {\bibfnamefont {Y.}~\bibnamefont {Singhal}},
  \bibinfo {author} {\bibfnamefont {B.}~\bibnamefont {Gadway}}, \bibinfo
  {author} {\bibfnamefont {T.}~\bibnamefont {Ozawa}}, \ and\ \bibinfo {author}
  {\bibfnamefont {H.~M.}\ \bibnamefont {Price}},\ }\href {\doibase
  10.1103/PhysRevE.107.064211} {\bibfield  {journal} {\bibinfo  {journal}
  {Phys. Rev. E}\ }\textbf {\bibinfo {volume} {107}},\ \bibinfo {pages}
  {064211} (\bibinfo {year} {2023})}\BibitemShut {NoStop}%
\bibitem [{\citenamefont {Singhal}\ \emph {et~al.}(2023)\citenamefont
  {Singhal}, \citenamefont {Martello}, \citenamefont {Agrawal}, \citenamefont
  {Ozawa}, \citenamefont {Price},\ and\ \citenamefont
  {Gadway}}]{Singhal-NH-AB}%
  \BibitemOpen
  \bibfield  {author} {\bibinfo {author} {\bibfnamefont {Y.}~\bibnamefont
  {Singhal}}, \bibinfo {author} {\bibfnamefont {E.}~\bibnamefont {Martello}},
  \bibinfo {author} {\bibfnamefont {S.}~\bibnamefont {Agrawal}}, \bibinfo
  {author} {\bibfnamefont {T.}~\bibnamefont {Ozawa}}, \bibinfo {author}
  {\bibfnamefont {H.}~\bibnamefont {Price}}, \ and\ \bibinfo {author}
  {\bibfnamefont {B.}~\bibnamefont {Gadway}},\ }\href {\doibase
  10.1103/PhysRevResearch.5.L032026} {\bibfield  {journal} {\bibinfo  {journal}
  {Phys. Rev. Res.}\ }\textbf {\bibinfo {volume} {5}},\ \bibinfo {pages}
  {L032026} (\bibinfo {year} {2023})}\BibitemShut {NoStop}%
\bibitem [{\citenamefont {Tian}\ \emph {et~al.}(2024)\citenamefont {Tian},
  \citenamefont {Velkovsky}, \citenamefont {Chen}, \citenamefont {Sun},
  \citenamefont {He},\ and\ \citenamefont {Gadway}}]{tian2023observation}%
  \BibitemOpen
  \bibfield  {author} {\bibinfo {author} {\bibfnamefont {M.}~\bibnamefont
  {Tian}}, \bibinfo {author} {\bibfnamefont {I.}~\bibnamefont {Velkovsky}},
  \bibinfo {author} {\bibfnamefont {T.}~\bibnamefont {Chen}}, \bibinfo {author}
  {\bibfnamefont {F.}~\bibnamefont {Sun}}, \bibinfo {author} {\bibfnamefont
  {Q.}~\bibnamefont {He}}, \ and\ \bibinfo {author} {\bibfnamefont
  {B.}~\bibnamefont {Gadway}},\ }\href {\doibase
  10.1103/PhysRevLett.132.126602} {\bibfield  {journal} {\bibinfo  {journal}
  {Phys. Rev. Lett.}\ }\textbf {\bibinfo {volume} {132}},\ \bibinfo {pages}
  {126602} (\bibinfo {year} {2024})}\BibitemShut {NoStop}%
\bibitem [{Sup()}]{SuppMats}%
  \BibitemOpen
  \href@noop {} {}\bibinfo {note} {See Supplementary Material for more details
  on the experimental system, the analysis of the data, and the theoretical
  analysis.}\BibitemShut {Stop}%
\bibitem [{\citenamefont {Roushan}\ \emph {et~al.}(2017)\citenamefont
  {Roushan}, \citenamefont {Neill}, \citenamefont {Megrant}, \citenamefont
  {Chen}, \citenamefont {Babbush}, \citenamefont {Barends}, \citenamefont
  {Campbell}, \citenamefont {Chen}, \citenamefont {Chiaro}, \citenamefont
  {Dunsworth}, \citenamefont {Fowler}, \citenamefont {Jeffrey}, \citenamefont
  {Kelly}, \citenamefont {Lucero}, \citenamefont {Mutus}, \citenamefont
  {O'Malley}, \citenamefont {Neeley}, \citenamefont {Quintana}, \citenamefont
  {Sank}, \citenamefont {Vainsencher}, \citenamefont {Wenner}, \citenamefont
  {White}, \citenamefont {Kapit}, \citenamefont {Neven},\ and\ \citenamefont
  {Martinis}}]{Roushan2017}%
  \BibitemOpen
  \bibfield  {author} {\bibinfo {author} {\bibfnamefont {P.}~\bibnamefont
  {Roushan}}, \bibinfo {author} {\bibfnamefont {C.}~\bibnamefont {Neill}},
  \bibinfo {author} {\bibfnamefont {A.}~\bibnamefont {Megrant}}, \bibinfo
  {author} {\bibfnamefont {Y.}~\bibnamefont {Chen}}, \bibinfo {author}
  {\bibfnamefont {R.}~\bibnamefont {Babbush}}, \bibinfo {author} {\bibfnamefont
  {R.}~\bibnamefont {Barends}}, \bibinfo {author} {\bibfnamefont
  {B.}~\bibnamefont {Campbell}}, \bibinfo {author} {\bibfnamefont
  {Z.}~\bibnamefont {Chen}}, \bibinfo {author} {\bibfnamefont {B.}~\bibnamefont
  {Chiaro}}, \bibinfo {author} {\bibfnamefont {A.}~\bibnamefont {Dunsworth}},
  \bibinfo {author} {\bibfnamefont {A.}~\bibnamefont {Fowler}}, \bibinfo
  {author} {\bibfnamefont {E.}~\bibnamefont {Jeffrey}}, \bibinfo {author}
  {\bibfnamefont {J.}~\bibnamefont {Kelly}}, \bibinfo {author} {\bibfnamefont
  {E.}~\bibnamefont {Lucero}}, \bibinfo {author} {\bibfnamefont
  {J.}~\bibnamefont {Mutus}}, \bibinfo {author} {\bibfnamefont {P.~J.~J.}\
  \bibnamefont {O'Malley}}, \bibinfo {author} {\bibfnamefont {M.}~\bibnamefont
  {Neeley}}, \bibinfo {author} {\bibfnamefont {C.}~\bibnamefont {Quintana}},
  \bibinfo {author} {\bibfnamefont {D.}~\bibnamefont {Sank}}, \bibinfo {author}
  {\bibfnamefont {A.}~\bibnamefont {Vainsencher}}, \bibinfo {author}
  {\bibfnamefont {J.}~\bibnamefont {Wenner}}, \bibinfo {author} {\bibfnamefont
  {T.}~\bibnamefont {White}}, \bibinfo {author} {\bibfnamefont
  {E.}~\bibnamefont {Kapit}}, \bibinfo {author} {\bibfnamefont
  {H.}~\bibnamefont {Neven}}, \ and\ \bibinfo {author} {\bibfnamefont
  {J.}~\bibnamefont {Martinis}},\ }\href {\doibase 10.1038/nphys3930}
  {\bibfield  {journal} {\bibinfo  {journal} {Nature Physics}\ }\textbf
  {\bibinfo {volume} {13}},\ \bibinfo {pages} {146} (\bibinfo {year}
  {2017})}\BibitemShut {NoStop}%
\bibitem [{\citenamefont {Ozawa}\ \emph {et~al.}(2019)\citenamefont {Ozawa},
  \citenamefont {Price}, \citenamefont {Amo}, \citenamefont {Goldman},
  \citenamefont {Hafezi}, \citenamefont {Lu}, \citenamefont {Rechtsman},
  \citenamefont {Schuster}, \citenamefont {Simon}, \citenamefont {Zilberberg},\
  and\ \citenamefont {Carusotto}}]{top-phot}%
  \BibitemOpen
  \bibfield  {author} {\bibinfo {author} {\bibfnamefont {T.}~\bibnamefont
  {Ozawa}}, \bibinfo {author} {\bibfnamefont {H.~M.}\ \bibnamefont {Price}},
  \bibinfo {author} {\bibfnamefont {A.}~\bibnamefont {Amo}}, \bibinfo {author}
  {\bibfnamefont {N.}~\bibnamefont {Goldman}}, \bibinfo {author} {\bibfnamefont
  {M.}~\bibnamefont {Hafezi}}, \bibinfo {author} {\bibfnamefont
  {L.}~\bibnamefont {Lu}}, \bibinfo {author} {\bibfnamefont {M.~C.}\
  \bibnamefont {Rechtsman}}, \bibinfo {author} {\bibfnamefont {D.}~\bibnamefont
  {Schuster}}, \bibinfo {author} {\bibfnamefont {J.}~\bibnamefont {Simon}},
  \bibinfo {author} {\bibfnamefont {O.}~\bibnamefont {Zilberberg}}, \ and\
  \bibinfo {author} {\bibfnamefont {I.}~\bibnamefont {Carusotto}},\ }\href
  {\doibase 10.1103/RevModPhys.91.015006} {\bibfield  {journal} {\bibinfo
  {journal} {Rev. Mod. Phys.}\ }\textbf {\bibinfo {volume} {91}},\ \bibinfo
  {pages} {015006} (\bibinfo {year} {2019})}\BibitemShut {NoStop}%
\bibitem [{\citenamefont {Cooper}\ \emph {et~al.}(2019)\citenamefont {Cooper},
  \citenamefont {Dalibard},\ and\ \citenamefont
  {Spielman}}]{TopBands-Spielman}%
  \BibitemOpen
  \bibfield  {author} {\bibinfo {author} {\bibfnamefont {N.~R.}\ \bibnamefont
  {Cooper}}, \bibinfo {author} {\bibfnamefont {J.}~\bibnamefont {Dalibard}}, \
  and\ \bibinfo {author} {\bibfnamefont {I.~B.}\ \bibnamefont {Spielman}},\
  }\href {\doibase 10.1103/RevModPhys.91.015005} {\bibfield  {journal}
  {\bibinfo  {journal} {Rev. Mod. Phys.}\ }\textbf {\bibinfo {volume} {91}},\
  \bibinfo {pages} {015005} (\bibinfo {year} {2019})}\BibitemShut {NoStop}%
\bibitem [{\citenamefont {Salerno}\ \emph {et~al.}(2016)\citenamefont
  {Salerno}, \citenamefont {Ozawa}, \citenamefont {Price},\ and\ \citenamefont
  {Carusotto}}]{salerno2016}%
  \BibitemOpen
  \bibfield  {author} {\bibinfo {author} {\bibfnamefont {G.}~\bibnamefont
  {Salerno}}, \bibinfo {author} {\bibfnamefont {T.}~\bibnamefont {Ozawa}},
  \bibinfo {author} {\bibfnamefont {H.~M.}\ \bibnamefont {Price}}, \ and\
  \bibinfo {author} {\bibfnamefont {I.}~\bibnamefont {Carusotto}},\ }\href
  {\doibase 10.1103/PhysRevB.93.085105} {\bibfield  {journal} {\bibinfo
  {journal} {Phys. Rev. B}\ }\textbf {\bibinfo {volume} {93}},\ \bibinfo
  {pages} {085105} (\bibinfo {year} {2016})}\BibitemShut {NoStop}%
\bibitem [{\citenamefont {Nash}\ \emph {et~al.}(2015)\citenamefont {Nash},
  \citenamefont {Kleckner}, \citenamefont {Read}, \citenamefont {Vitelli},
  \citenamefont {Turner},\ and\ \citenamefont
  {Irvine}}]{GyroscopeMetamaterials}%
  \BibitemOpen
  \bibfield  {author} {\bibinfo {author} {\bibfnamefont {L.~M.}\ \bibnamefont
  {Nash}}, \bibinfo {author} {\bibfnamefont {D.}~\bibnamefont {Kleckner}},
  \bibinfo {author} {\bibfnamefont {A.}~\bibnamefont {Read}}, \bibinfo {author}
  {\bibfnamefont {V.}~\bibnamefont {Vitelli}}, \bibinfo {author} {\bibfnamefont
  {A.~M.}\ \bibnamefont {Turner}}, \ and\ \bibinfo {author} {\bibfnamefont
  {W.~T.~M.}\ \bibnamefont {Irvine}},\ }\href {\doibase
  10.1073/pnas.1507413112} {\bibfield  {journal} {\bibinfo  {journal}
  {Proceedings of the National Academy of Sciences}\ }\textbf {\bibinfo
  {volume} {112}},\ \bibinfo {pages} {14495} (\bibinfo {year}
  {2015})}\BibitemShut {NoStop}%
\bibitem [{\citenamefont {Süsstrunk}\ and\ \citenamefont
  {Huber}(2015)}]{HuberPendulum}%
  \BibitemOpen
  \bibfield  {author} {\bibinfo {author} {\bibfnamefont {R.}~\bibnamefont
  {Süsstrunk}}\ and\ \bibinfo {author} {\bibfnamefont {S.~D.}\ \bibnamefont
  {Huber}},\ }\href {\doibase 10.1126/science.aab0239} {\bibfield  {journal}
  {\bibinfo  {journal} {Science}\ }\textbf {\bibinfo {volume} {349}},\ \bibinfo
  {pages} {47} (\bibinfo {year} {2015})}\BibitemShut {NoStop}%
\bibitem [{\citenamefont {S{\"u}sstrunk}\ and\ \citenamefont
  {Huber}(2016)}]{susstrunk2016classification}%
  \BibitemOpen
  \bibfield  {author} {\bibinfo {author} {\bibfnamefont {R.}~\bibnamefont
  {S{\"u}sstrunk}}\ and\ \bibinfo {author} {\bibfnamefont {S.~D.}\ \bibnamefont
  {Huber}},\ }\href {\doibase 10.1073/pnas.1605462113} {\bibfield  {journal}
  {\bibinfo  {journal} {Proceedings of the National Academy of Sciences}\
  }\textbf {\bibinfo {volume} {113}},\ \bibinfo {pages} {E4767} (\bibinfo
  {year} {2016})}\BibitemShut {NoStop}%
\bibitem [{\citenamefont {Huber}(2016)}]{Huber2016}%
  \BibitemOpen
  \bibfield  {author} {\bibinfo {author} {\bibfnamefont {S.~D.}\ \bibnamefont
  {Huber}},\ }\href {\doibase 10.1038/nphys3801} {\bibfield  {journal}
  {\bibinfo  {journal} {Nature Physics}\ }\textbf {\bibinfo {volume} {12}},\
  \bibinfo {pages} {621} (\bibinfo {year} {2016})}\BibitemShut {NoStop}%
\bibitem [{\citenamefont {Mathew}\ \emph {et~al.}(2020)\citenamefont {Mathew},
  \citenamefont {Pino},\ and\ \citenamefont {Verhagen}}]{Phonon-gaugefield}%
  \BibitemOpen
  \bibfield  {author} {\bibinfo {author} {\bibfnamefont {J.~P.}\ \bibnamefont
  {Mathew}}, \bibinfo {author} {\bibfnamefont {J.~d.}\ \bibnamefont {Pino}}, \
  and\ \bibinfo {author} {\bibfnamefont {E.}~\bibnamefont {Verhagen}},\ }\href
  {\doibase 10.1038/s41565-019-0630-8} {\bibfield  {journal} {\bibinfo
  {journal} {Nature Nanotechnology}\ }\textbf {\bibinfo {volume} {15}},\
  \bibinfo {pages} {198} (\bibinfo {year} {2020})}\BibitemShut {NoStop}%
\bibitem [{\citenamefont {Albiez}\ \emph {et~al.}(2005)\citenamefont {Albiez},
  \citenamefont {Gati}, \citenamefont {F\"olling}, \citenamefont {Hunsmann},
  \citenamefont {Cristiani},\ and\ \citenamefont
  {Oberthaler}}]{Albiez-DirectJosephson}%
  \BibitemOpen
  \bibfield  {author} {\bibinfo {author} {\bibfnamefont {M.}~\bibnamefont
  {Albiez}}, \bibinfo {author} {\bibfnamefont {R.}~\bibnamefont {Gati}},
  \bibinfo {author} {\bibfnamefont {J.}~\bibnamefont {F\"olling}}, \bibinfo
  {author} {\bibfnamefont {S.}~\bibnamefont {Hunsmann}}, \bibinfo {author}
  {\bibfnamefont {M.}~\bibnamefont {Cristiani}}, \ and\ \bibinfo {author}
  {\bibfnamefont {M.~K.}\ \bibnamefont {Oberthaler}},\ }\href {\doibase
  10.1103/PhysRevLett.95.010402} {\bibfield  {journal} {\bibinfo  {journal}
  {Phys. Rev. Lett.}\ }\textbf {\bibinfo {volume} {95}},\ \bibinfo {pages}
  {010402} (\bibinfo {year} {2005})}\BibitemShut {NoStop}%
\bibitem [{\citenamefont {Zibold}\ \emph {et~al.}(2010)\citenamefont {Zibold},
  \citenamefont {Nicklas}, \citenamefont {Gross},\ and\ \citenamefont
  {Oberthaler}}]{Zibold}%
  \BibitemOpen
  \bibfield  {author} {\bibinfo {author} {\bibfnamefont {T.}~\bibnamefont
  {Zibold}}, \bibinfo {author} {\bibfnamefont {E.}~\bibnamefont {Nicklas}},
  \bibinfo {author} {\bibfnamefont {C.}~\bibnamefont {Gross}}, \ and\ \bibinfo
  {author} {\bibfnamefont {M.~K.}\ \bibnamefont {Oberthaler}},\ }\href
  {\doibase 10.1103/PhysRevLett.105.204101} {\bibfield  {journal} {\bibinfo
  {journal} {Phys. Rev. Lett.}\ }\textbf {\bibinfo {volume} {105}},\ \bibinfo
  {pages} {204101} (\bibinfo {year} {2010})}\BibitemShut {NoStop}%
\bibitem [{\citenamefont {Sacha}\ and\ \citenamefont
  {Zakrzewski}(2017)}]{Sacha_2018}%
  \BibitemOpen
  \bibfield  {author} {\bibinfo {author} {\bibfnamefont {K.}~\bibnamefont
  {Sacha}}\ and\ \bibinfo {author} {\bibfnamefont {J.}~\bibnamefont
  {Zakrzewski}},\ }\href {\doibase 10.1088/1361-6633/aa8b38} {\bibfield
  {journal} {\bibinfo  {journal} {Reports on Progress in Physics}\ }\textbf
  {\bibinfo {volume} {81}},\ \bibinfo {pages} {016401} (\bibinfo {year}
  {2017})}\BibitemShut {NoStop}%
\bibitem [{\citenamefont {Heugel}\ \emph {et~al.}(2019)\citenamefont {Heugel},
  \citenamefont {Oscity}, \citenamefont {Eichler}, \citenamefont {Zilberberg},\
  and\ \citenamefont {Chitra}}]{Oded-TC}%
  \BibitemOpen
  \bibfield  {author} {\bibinfo {author} {\bibfnamefont {T.~L.}\ \bibnamefont
  {Heugel}}, \bibinfo {author} {\bibfnamefont {M.}~\bibnamefont {Oscity}},
  \bibinfo {author} {\bibfnamefont {A.}~\bibnamefont {Eichler}}, \bibinfo
  {author} {\bibfnamefont {O.}~\bibnamefont {Zilberberg}}, \ and\ \bibinfo
  {author} {\bibfnamefont {R.}~\bibnamefont {Chitra}},\ }\href {\doibase
  10.1103/PhysRevLett.123.124301} {\bibfield  {journal} {\bibinfo  {journal}
  {Phys. Rev. Lett.}\ }\textbf {\bibinfo {volume} {123}},\ \bibinfo {pages}
  {124301} (\bibinfo {year} {2019})}\BibitemShut {NoStop}%
\bibitem [{\citenamefont {Yao}\ \emph {et~al.}(2020)\citenamefont {Yao},
  \citenamefont {Nayak}, \citenamefont {Balents},\ and\ \citenamefont
  {Zaletel}}]{yao-np2020-timecrystals}%
  \BibitemOpen
  \bibfield  {author} {\bibinfo {author} {\bibfnamefont {N.~Y.}\ \bibnamefont
  {Yao}}, \bibinfo {author} {\bibfnamefont {C.}~\bibnamefont {Nayak}}, \bibinfo
  {author} {\bibfnamefont {L.}~\bibnamefont {Balents}}, \ and\ \bibinfo
  {author} {\bibfnamefont {M.~P.}\ \bibnamefont {Zaletel}},\ }\href {\doibase
  10.1038/s41567-019-0782-3} {\bibfield  {journal} {\bibinfo  {journal} {Nat.
  Phys.}\ }\textbf {\bibinfo {volume} {16}},\ \bibinfo {pages} {438} (\bibinfo
  {year} {2020})}\BibitemShut {NoStop}%
\bibitem [{\citenamefont {Zaletel}\ \emph {et~al.}(2023)\citenamefont
  {Zaletel}, \citenamefont {Lukin}, \citenamefont {Monroe}, \citenamefont
  {Nayak}, \citenamefont {Wilczek},\ and\ \citenamefont
  {Yao}}]{timecrystal-RMP}%
  \BibitemOpen
  \bibfield  {author} {\bibinfo {author} {\bibfnamefont {M.~P.}\ \bibnamefont
  {Zaletel}}, \bibinfo {author} {\bibfnamefont {M.}~\bibnamefont {Lukin}},
  \bibinfo {author} {\bibfnamefont {C.}~\bibnamefont {Monroe}}, \bibinfo
  {author} {\bibfnamefont {C.}~\bibnamefont {Nayak}}, \bibinfo {author}
  {\bibfnamefont {F.}~\bibnamefont {Wilczek}}, \ and\ \bibinfo {author}
  {\bibfnamefont {N.~Y.}\ \bibnamefont {Yao}},\ }\href {\doibase
  10.1103/RevModPhys.95.031001} {\bibfield  {journal} {\bibinfo  {journal}
  {Rev. Mod. Phys.}\ }\textbf {\bibinfo {volume} {95}},\ \bibinfo {pages}
  {031001} (\bibinfo {year} {2023})}\BibitemShut {NoStop}%
\bibitem [{\citenamefont {Hannaford}\ and\ \citenamefont
  {Sacha}(2022)}]{Hannaford_2022}%
  \BibitemOpen
  \bibfield  {author} {\bibinfo {author} {\bibfnamefont {P.}~\bibnamefont
  {Hannaford}}\ and\ \bibinfo {author} {\bibfnamefont {K.}~\bibnamefont
  {Sacha}},\ }\href {\doibase 10.1209/0295-5075/ac796d} {\bibfield  {journal}
  {\bibinfo  {journal} {Europhysics Letters}\ }\textbf {\bibinfo {volume}
  {139}},\ \bibinfo {pages} {10001} (\bibinfo {year} {2022})}\BibitemShut
  {NoStop}%
\bibitem [{\citenamefont {Avni}\ \emph {et~al.}(2023)\citenamefont {Avni},
  \citenamefont {Fruchart}, \citenamefont {Martin}, \citenamefont {Seara},\
  and\ \citenamefont {Vitelli}}]{Vitelli-Ising}%
  \BibitemOpen
  \bibfield  {author} {\bibinfo {author} {\bibfnamefont {Y.}~\bibnamefont
  {Avni}}, \bibinfo {author} {\bibfnamefont {M.}~\bibnamefont {Fruchart}},
  \bibinfo {author} {\bibfnamefont {D.}~\bibnamefont {Martin}}, \bibinfo
  {author} {\bibfnamefont {D.}~\bibnamefont {Seara}}, \ and\ \bibinfo {author}
  {\bibfnamefont {V.}~\bibnamefont {Vitelli}},\ }\href@noop {} {\enquote
  {\bibinfo {title} {The non-reciprocal ising model},}\ } (\bibinfo {year}
  {2023}),\ \Eprint {http://arxiv.org/abs/2311.05471} {arXiv:2311.05471
  [cond-mat.stat-mech]} \BibitemShut {NoStop}%
\bibitem [{\citenamefont {Liu}\ \emph {et~al.}(2023)\citenamefont {Liu},
  \citenamefont {Ou}, \citenamefont {MacDonald},\ and\ \citenamefont
  {Zheludev}}]{time-crystal-nonrecip}%
  \BibitemOpen
  \bibfield  {author} {\bibinfo {author} {\bibfnamefont {T.}~\bibnamefont
  {Liu}}, \bibinfo {author} {\bibfnamefont {J.-Y.}\ \bibnamefont {Ou}},
  \bibinfo {author} {\bibfnamefont {K.~F.}\ \bibnamefont {MacDonald}}, \ and\
  \bibinfo {author} {\bibfnamefont {N.~I.}\ \bibnamefont {Zheludev}},\ }\href
  {\doibase 10.1038/s41567-023-02023-5} {\bibfield  {journal} {\bibinfo
  {journal} {Nature Physics}\ }\textbf {\bibinfo {volume} {19}},\ \bibinfo
  {pages} {986} (\bibinfo {year} {2023})}\BibitemShut {NoStop}%
\bibitem [{\citenamefont {Hanai}(2024)}]{Time-crystal-nonr}%
  \BibitemOpen
  \bibfield  {author} {\bibinfo {author} {\bibfnamefont {R.}~\bibnamefont
  {Hanai}},\ }\href {\doibase 10.1103/PhysRevX.14.011029} {\bibfield  {journal}
  {\bibinfo  {journal} {Phys. Rev. X}\ }\textbf {\bibinfo {volume} {14}},\
  \bibinfo {pages} {011029} (\bibinfo {year} {2024})}\BibitemShut {NoStop}%
\bibitem [{\citenamefont {Kozin}\ and\ \citenamefont
  {Kyriienko}(2019)}]{Kozin-LR}%
  \BibitemOpen
  \bibfield  {author} {\bibinfo {author} {\bibfnamefont {V.~K.}\ \bibnamefont
  {Kozin}}\ and\ \bibinfo {author} {\bibfnamefont {O.}~\bibnamefont
  {Kyriienko}},\ }\href {\doibase 10.1103/PhysRevLett.123.210602} {\bibfield
  {journal} {\bibinfo  {journal} {Phys. Rev. Lett.}\ }\textbf {\bibinfo
  {volume} {123}},\ \bibinfo {pages} {210602} (\bibinfo {year}
  {2019})}\BibitemShut {NoStop}%
\bibitem [{\citenamefont {Ke\ss{}ler}\ \emph {et~al.}(2021)\citenamefont
  {Ke\ss{}ler}, \citenamefont {Kongkhambut}, \citenamefont {Georges},
  \citenamefont {Mathey}, \citenamefont {Cosme},\ and\ \citenamefont
  {Hemmerich}}]{Hemmerich-diss-TC}%
  \BibitemOpen
  \bibfield  {author} {\bibinfo {author} {\bibfnamefont {H.}~\bibnamefont
  {Ke\ss{}ler}}, \bibinfo {author} {\bibfnamefont {P.}~\bibnamefont
  {Kongkhambut}}, \bibinfo {author} {\bibfnamefont {C.}~\bibnamefont
  {Georges}}, \bibinfo {author} {\bibfnamefont {L.}~\bibnamefont {Mathey}},
  \bibinfo {author} {\bibfnamefont {J.~G.}\ \bibnamefont {Cosme}}, \ and\
  \bibinfo {author} {\bibfnamefont {A.}~\bibnamefont {Hemmerich}},\ }\href
  {\doibase 10.1103/PhysRevLett.127.043602} {\bibfield  {journal} {\bibinfo
  {journal} {Phys. Rev. Lett.}\ }\textbf {\bibinfo {volume} {127}},\ \bibinfo
  {pages} {043602} (\bibinfo {year} {2021})}\BibitemShut {NoStop}%
\bibitem [{\citenamefont {Greschner}\ \emph {et~al.}(2015)\citenamefont
  {Greschner}, \citenamefont {Piraud}, \citenamefont {Heidrich-Meisner},
  \citenamefont {McCulloch}, \citenamefont {Schollw\"ock},\ and\ \citenamefont
  {Vekua}}]{Vekua-reversal}%
  \BibitemOpen
  \bibfield  {author} {\bibinfo {author} {\bibfnamefont {S.}~\bibnamefont
  {Greschner}}, \bibinfo {author} {\bibfnamefont {M.}~\bibnamefont {Piraud}},
  \bibinfo {author} {\bibfnamefont {F.}~\bibnamefont {Heidrich-Meisner}},
  \bibinfo {author} {\bibfnamefont {I.~P.}\ \bibnamefont {McCulloch}}, \bibinfo
  {author} {\bibfnamefont {U.}~\bibnamefont {Schollw\"ock}}, \ and\ \bibinfo
  {author} {\bibfnamefont {T.}~\bibnamefont {Vekua}},\ }\href {\doibase
  10.1103/PhysRevLett.115.190402} {\bibfield  {journal} {\bibinfo  {journal}
  {Phys. Rev. Lett.}\ }\textbf {\bibinfo {volume} {115}},\ \bibinfo {pages}
  {190402} (\bibinfo {year} {2015})}\BibitemShut {NoStop}%
\bibitem [{\citenamefont {Buser}\ \emph {et~al.}(2020)\citenamefont {Buser},
  \citenamefont {Hubig}, \citenamefont {Schollw\"ock}, \citenamefont
  {Tarruell},\ and\ \citenamefont {Heidrich-Meisner}}]{synth-flux-ladder}%
  \BibitemOpen
  \bibfield  {author} {\bibinfo {author} {\bibfnamefont {M.}~\bibnamefont
  {Buser}}, \bibinfo {author} {\bibfnamefont {C.}~\bibnamefont {Hubig}},
  \bibinfo {author} {\bibfnamefont {U.}~\bibnamefont {Schollw\"ock}}, \bibinfo
  {author} {\bibfnamefont {L.}~\bibnamefont {Tarruell}}, \ and\ \bibinfo
  {author} {\bibfnamefont {F.}~\bibnamefont {Heidrich-Meisner}},\ }\href
  {\doibase 10.1103/PhysRevA.102.053314} {\bibfield  {journal} {\bibinfo
  {journal} {Phys. Rev. A}\ }\textbf {\bibinfo {volume} {102}},\ \bibinfo
  {pages} {053314} (\bibinfo {year} {2020})}\BibitemShut {NoStop}%
\bibitem [{\citenamefont {Edmonds}\ \emph {et~al.}(2013)\citenamefont
  {Edmonds}, \citenamefont {Valiente}, \citenamefont
  {Juzeli\ifmmode~\bar{u}\else \={u}\fi{}nas}, \citenamefont {Santos},\ and\
  \citenamefont {\"Ohberg}}]{Edmonds-Gauge}%
  \BibitemOpen
  \bibfield  {author} {\bibinfo {author} {\bibfnamefont {M.~J.}\ \bibnamefont
  {Edmonds}}, \bibinfo {author} {\bibfnamefont {M.}~\bibnamefont {Valiente}},
  \bibinfo {author} {\bibfnamefont {G.}~\bibnamefont
  {Juzeli\ifmmode~\bar{u}\else \={u}\fi{}nas}}, \bibinfo {author}
  {\bibfnamefont {L.}~\bibnamefont {Santos}}, \ and\ \bibinfo {author}
  {\bibfnamefont {P.}~\bibnamefont {\"Ohberg}},\ }\href {\doibase
  10.1103/PhysRevLett.110.085301} {\bibfield  {journal} {\bibinfo  {journal}
  {Phys. Rev. Lett.}\ }\textbf {\bibinfo {volume} {110}},\ \bibinfo {pages}
  {085301} (\bibinfo {year} {2013})}\BibitemShut {NoStop}%
\bibitem [{\citenamefont {Fr{\"o}lian}\ \emph {et~al.}(2022)\citenamefont
  {Fr{\"o}lian}, \citenamefont {Chisholm}, \citenamefont {Neri}, \citenamefont
  {Cabrera}, \citenamefont {Ramos}, \citenamefont {Celi},\ and\ \citenamefont
  {Tarruell}}]{Frolian2022}%
  \BibitemOpen
  \bibfield  {author} {\bibinfo {author} {\bibfnamefont {A.}~\bibnamefont
  {Fr{\"o}lian}}, \bibinfo {author} {\bibfnamefont {C.~S.}\ \bibnamefont
  {Chisholm}}, \bibinfo {author} {\bibfnamefont {E.}~\bibnamefont {Neri}},
  \bibinfo {author} {\bibfnamefont {C.~R.}\ \bibnamefont {Cabrera}}, \bibinfo
  {author} {\bibfnamefont {R.}~\bibnamefont {Ramos}}, \bibinfo {author}
  {\bibfnamefont {A.}~\bibnamefont {Celi}}, \ and\ \bibinfo {author}
  {\bibfnamefont {L.}~\bibnamefont {Tarruell}},\ }\href {\doibase
  10.1038/s41586-022-04943-3} {\bibfield  {journal} {\bibinfo  {journal}
  {Nature}\ }\textbf {\bibinfo {volume} {608}},\ \bibinfo {pages} {293}
  (\bibinfo {year} {2022})}\BibitemShut {NoStop}%
\bibitem [{\citenamefont {Faugno}\ \emph {et~al.}(2024)\citenamefont {Faugno},
  \citenamefont {Salerno},\ and\ \citenamefont {Ozawa}}]{Faugno-dens}%
  \BibitemOpen
  \bibfield  {author} {\bibinfo {author} {\bibfnamefont {W.~N.}\ \bibnamefont
  {Faugno}}, \bibinfo {author} {\bibfnamefont {M.}~\bibnamefont {Salerno}}, \
  and\ \bibinfo {author} {\bibfnamefont {T.}~\bibnamefont {Ozawa}},\ }\href
  {\doibase 10.1103/PhysRevLett.132.023401} {\bibfield  {journal} {\bibinfo
  {journal} {Phys. Rev. Lett.}\ }\textbf {\bibinfo {volume} {132}},\ \bibinfo
  {pages} {023401} (\bibinfo {year} {2024})}\BibitemShut {NoStop}%
\bibitem [{\citenamefont {Riechert}\ \emph {et~al.}(2022)\citenamefont
  {Riechert}, \citenamefont {Halimeh}, \citenamefont {Kasper}, \citenamefont
  {Bretheau}, \citenamefont {Zohar}, \citenamefont {Hauke},\ and\ \citenamefont
  {Jendrzejewski}}]{FredJlgt}%
  \BibitemOpen
  \bibfield  {author} {\bibinfo {author} {\bibfnamefont {H.}~\bibnamefont
  {Riechert}}, \bibinfo {author} {\bibfnamefont {J.~C.}\ \bibnamefont
  {Halimeh}}, \bibinfo {author} {\bibfnamefont {V.}~\bibnamefont {Kasper}},
  \bibinfo {author} {\bibfnamefont {L.}~\bibnamefont {Bretheau}}, \bibinfo
  {author} {\bibfnamefont {E.}~\bibnamefont {Zohar}}, \bibinfo {author}
  {\bibfnamefont {P.}~\bibnamefont {Hauke}}, \ and\ \bibinfo {author}
  {\bibfnamefont {F.}~\bibnamefont {Jendrzejewski}},\ }\href {\doibase
  10.1103/PhysRevB.105.205141} {\bibfield  {journal} {\bibinfo  {journal}
  {Phys. Rev. B}\ }\textbf {\bibinfo {volume} {105}},\ \bibinfo {pages}
  {205141} (\bibinfo {year} {2022})}\BibitemShut {NoStop}%
\end{thebibliography}%

\end{document}


\begin{widetext}
\appendix

\section{Supplemental Material for \\ ``Observation of chiral solitary waves in a nonlinear Aharonov-Bohm ring''}

\vspace{5mm}
\section{Theoretical derivation}
Our discrete Josephson ring realized in an active mechanical network of three oscillators is based on mapping Newton's equations of motion onto the Heisenberg equations for a desired target Hamiltonian. 
A detailed treatment of this derivation is provided in Refs.~\cite{Anandwade-synthetic,tian2023observation}.
We begin with the equations of motion of a set of identical classical harmonic oscillators with angular frequency $\omega$ and mass $m$:
\begin{align}
m \dot{x}_i(t) = p_i(t),\ \ \ \dot{p}_i(t) = -m\omega^2 x_i(t) \ ,
\label{eq:NewtonOsc}
\end{align}
where $(x_i, \  p_i)$ are the position and momentum of the $i$-th oscillator, respectively. For simplicity of notation, we introduce the variables $\tilde{x}_i \equiv -\omega^2 x_i(t)$ and $\tilde{p}_i \equiv (-\omega^2 / m) p_i(t)$
and rewrite the equations of motion as:
\begin{align}
\dot{\tilde{x}}_i = \tilde{p}_i,\ \ \ \dot{\tilde{p}}_i = -\omega^2 \tilde{x}_i \ .
\label{eq:NewtonOsc2}
\end{align}

To couple the oscillators as a basis for simulating different model Hamiltonians, we add feedback forces such that the equations of motion become:
\begin{align}
\dot{\tilde{x}}_i = \tilde{p}_i,\ \ \ \dot{\tilde{p}}_i = -\omega^2 \tilde{x}_i + f_i \ ,
\label{eq:NewtonOsc3}
\end{align}
where $f_i$ is a function of $(\tilde{x}_i,\ \tilde{p}_i)$. In order to map this system to the Heisenberg equations of motion, we introduce the classical complex amplitudes:
\begin{align}
\alpha_i \equiv \sqrt{\frac{\omega}{2}}\tilde{x}_i + i\sqrt{\frac{1}{2\omega}}\tilde{p}_i \ ,
\label{eq:compAmps}
\end{align}
which are analogous to the annihilation operator of the quantum harmonic oscillator. It follows that  
\begin{align}
\tilde{x}_i = \sqrt{\frac{1}{2\omega}}(\alpha_i + \alpha_i^*),\ \ \ \tilde{p}_i = -i\sqrt{\frac{\omega}{2}}(\alpha_i - \alpha_i^*)
\label{eq:XandP}
\end{align}
and hence the equations of motion become
\begin{align}
\dot{\alpha}_i = -i\omega\alpha_i + \frac{i}{\sqrt{2\omega}}f_i \ ,
\label{eq:eomAlpha}
\end{align}
with $f_i$ also expressed in terms of (being a function of) $(\alpha_i, \ \alpha_i^*)$.

In the absence of feedback, the complex amplitudes of the system evolve independently as $\alpha_i(t)\propto e^{i\omega t}$ (ignoring the natural dissipation of each oscillator, which we assume to be weak). The feedback forces will modify the dynamics. However, as long as they are sufficiently weak and $\omega$ remains the largest frequency-scale in the system, the changes will be small in comparison to the natural frequency of the oscillators. In this limit, we can apply the Rotating Wave Approximation (RWA) to re-write eq.~\ref{eq:eomAlpha} as
\begin{align}
\dot{\alpha}_i = -i\omega\alpha_i + \sum_j \frac{i}{\sqrt{2\omega}}F_{ij}^{\mathit{RWA}}\alpha_j \ ,
\label{eq:RWA-1}
\end{align}
where $\sum_jF_{ij}^{\mathit{RWA}}\alpha_j$ is obtained via applying the RWA to the applied feedback force (i.e., keeping only co-rotating terms). This can be regarded as the Heisenberg equation of motion of the bosonic tight-binding Hamiltonian (with $\hbar = 1$)
\begin{align}
\mathcal{H} = \sum_i\omega\hat{\alpha}_i^\dagger\hat{\alpha}_i + \frac{1}{\sqrt{2\omega}}\sum_{i,j} \frac{1}{n_i+1} \hat{\alpha}_i^\dagger \hat{F}_{ij}^{\mathit{RWA}}\hat{\alpha}_j \ .
\label{eq:H_TBM}
\end{align}
Here $(\hat{\alpha}^\dagger, \hat{\alpha})$ are bosonic creation and annihilation operators, and $\hat{F}_{ij}^{\mathit{RWA}}$ is obtained by replacing $(\alpha_i^*, \alpha_i)$ by their corresponding creation and annihilation operators in normal ordering.  
To obtain the forces that we should apply to our experimental system, we translate back to the $X$ and $P$ variables for each individual oscillator, keeping the real part of the force in the equations of motion.

\section{Experimental Hamiltonian}
The dynamics of this system can be described by the classical nonlinear wave equation ($\alpha \equiv \psi$ in main text)
\begin{align}
i \dot{\alpha}_j = (\omega - U |\alpha_j|^2)\alpha_j -J \sum_{k\neq j} e^{i \varphi_{kj}}\alpha_k \ ,
\label{eq:Ham-wave}
\end{align}
where $f_0 = \omega/2\pi = 12.9916(2)$~Hz is the common oscillator frequency, $J/2\pi$ is the effective phonon hopping rate, and $\varphi_{23} \equiv \varphi$ (with $\varphi_{12} = \varphi_{31} = 0$) sets the flux in the loop. Finally, $U$ represents the tunable strength of the nonlinear self interactions. That is, when all of the energy resides at just one of the oscillators, its frequency is shifted down by $U/(2\pi)$ relative to the others.

The feedback forces used in implementing the Hamiltonian for our experiment follow two forms - one implementing the complex inter-oscillator hopping, and the other providing the nonlinear self-interaction. First, we consider the inter-oscillator hopping between site $j$ and $k$, given by $J e^{i \varphi_{kj}}\alpha_k$ in the equation of motion (with the phases defined to be antisymmetric, i.e., $\varphi_{jk} = -\varphi_{kj}$). Translating this to our experimental signals $X$ and $P$, we get $F_j \propto e^{i \varphi_{kj}}(X_k + i P_k)$. As we can only apply real forces to the oscillators, we take the real part of this, yielding $F_j \propto \cos{\varphi_{kj}}X_k + \sin{\varphi_{kj}}P_k$. Similarly, the nonlinear self-interaction is implemented by taking $F_j \propto  (X_j^2 + P_j^2)X_j$, again keeping only the real part of the force.

An intuitive picture for the necessary forces can be found by considering that the applied feedback forces effectively implement Hamilton's equations. That is, for a desired model Hamiltonian $\mathcal{H}$, one applies forces such that $F_j \equiv \mathrm{d}p_j / dt \propto - \partial \mathcal{H} / \partial x_j$~\cite{Anandwade-synthetic,tian2023observation}.

\section{Experimental Implementation}

In this experiment, our “lattice of synthetically coupled oscillators” consists of modular mechanical oscillators, as described in Ref.~\cite{tian2023observation}. These oscillators are constructed of a 3D printed mass attached to a 3D printed support cage via two metal springs. In the absence of applied feedback forces, these oscillators exhibit nearly identical natural oscillation frequencies, $\omega_0 / 2\pi \approx 13$~Hz. Each oscillator is equipped with an embedded analog accelerometer (EVAL-ADXL203). Real-time measurements of acceleration $a(t)$ are obtained by sending the signals to a common computer (via very high-gauge wire connections).

Additionally, by numerically differentiating the acquired signal, we obtain real-time measurements of the oscillators’ jerk, $j(t) \equiv \dot{a}(t)$. Because $a(t) \propto x(t)$ and $j(t) \propto p(t)$ for a harmonic oscillator, we treat the $a$ and $j$ signals as proxies for the position $x$ and momentum $p$. We additionally multiply (i.e., normalize) the signals to put them on a common scale, reflecting the equipartition of harmonic oscillators.

As described in the previous section, we use these $x$ and $p$ proxies to implement the experimental Hamiltonian under study. We sample the signal at 1~kHz (i.e. $\Delta t = 1$~ms) and calculate the driving force needed at each oscillator. This force is implemented magnetically, avoiding any additional mechanical contacts. Each oscillator has a dipole magnet attached to a central, cylindrical shaft. The dipole magnet is embedded in a pair of anti-Helmholtz coils. We control the current in these coils, which produces an axial magnetic field gradient that in turn creates a force on the oscillator. We note that we operate the current in a polar fashion, with only one direction of current flow allowed.

We use feedback forces at each oscillator in order to cancel damping and reduce frequency disorder. A $p_i$ dependent force acts to counteract the damping, while an $x_i$ dependent force shifts the frequency of the site. This allows us to increase the effective quality factor of each site, as well as ensure frequency uniformity across the oscillators.

Our primary data consists of the same real-time measurements of the $x$ and $p$ signals for each of the oscillators. From these signals, we obtain the observables given below.

\section{IPR and winding}

In our experiment, we characterize the dynamics of the system by two metrics - the Inverse Participation Ratio (IPR) and the center-of-mass winding. The IPR provides a metric for the localization of the excitation in the system, while the winding measures the degree of chiral behavior. We obtain the average IPR over an experimental run of length $T$, given by
\begin{align}
\mathit{\overline{IPR}} = \frac{1}{T}\int_0^T\sum_i |\psi_i(t)|^4 dt \ ,
\label{eq:IPR-1}
\end{align}
by taking the oscillator energy $X_i^2 + P_i^2 = |\psi|^2$ (as per Eq.~\ref{eq:XandP}). We then average over the experimental run with a sampling time $\Delta t$, giving us
\begin{align}
\mathit{\overline{IPR}} = \frac{1}{T}\sum_{n=0}^{T/\Delta t}\sum_i (X_i(n\Delta t)^2 + P_i(n\Delta t)^2)^2 \ .
\label{eq:IPR-2}
\end{align}

As the total energy of the system is conserved, we can reduce the amplitude of the three oscillators over time to a two-dimensional trajectory that lies within the boundaries of a triangle. This is a two-dimensional projection of the four-dimensional phase space, where the relevant phase differences between the oscillators are projected out.
In this two-dimensional space, each vertex represents all of the energy in the system residing at one site. We define the center of mass coordinates as ($c_x = |\psi_3|^2 - |\psi_1|^2, c_y = \frac{|\psi_2|^2}{\sqrt{2}})$, and place the origin (i.e. $|\psi_1|^2 = |\psi_2|^2 = |\psi_3|^2$) at $(0, \frac{\sqrt{2}}{3})$. In this coordinate system, our experiment is initialized in the state with coordinates $(0.5,0)$ (i.e., all of the energy in oscillator 3). In order to obtain the winding, we define the angle of any point within the triangle as
\begin{align}
\theta = \atantwo(c_y - \frac{\sqrt{2}}{3},c_x) \ ,
\label{eq:Traj-angle}
\end{align}
where $\atantwo(y,x)$ is the two-argument arctangent function on the Cartesian plane. We then define the winding via the accumulated angle, given by the line integral over a trajectory $C$, which we parameterize in terms of time $t$ as
\begin{align}
\mathcal{W}(t) =\int_C \frac{1}{2\pi}d\theta = \int_0^t \frac{1}{2\pi} \frac{d\theta(t')}{dt'}dt' \ .
\label{eq:Winding-number}
\end{align}
As the $\atantwo$ function only produces results in the range $(-\pi,\pi]$, we cannot simply take the difference $\theta(t) - \theta(0)$ and must instead accumulate the angle as a discrete sum given by
\begin{align}
\mathcal{W}(t) = \sum_{n=1}^{t/\Delta t} \frac{1}{2\pi} \theta(n\Delta t) - \theta((n-1)\Delta t) \ .
\label{eq:Winding-summation-1}
\end{align}

However, there is a small wrinkle in the calculation, namely that $\atantwo$ and thus the angle is not well-defined at the origin (i.e. when the three sites have equal energy) and discontinuous across the line $(y=0, x<0)$. In order to deal with this, we make the assumption that our trajectory will vary smoothly as it passes through the discontinuity. Taking the time that the trajectory crosses the discontinuity to be $T$, we assume that
\begin{align}
\theta(T+\Delta t) = \theta(T - \Delta t) \pm \pi + \epsilon \ ,
\label{eq:angle-thru-origin}
\end{align}
where $\epsilon$ is a sufficiently small variation. Therefore, we can exploit the fact that $\sin(x + 2n\pi) \approx x$ for small $x$ to say that, for sampling time fast compared to the dynamics, $ \sin(\theta(n\Delta t) - \theta((n-1)\Delta t)) \approx  \theta(n\Delta t) - \theta((n-1)\Delta t)$. Thus, our winding is calculated as
\begin{align}
\mathcal{W}(t) \approx \sum_{n=1}^{t/\Delta t} \frac{1}{2\pi} \sin(\theta(n\Delta t) - \theta((n-1)\Delta t)) \ .
\label{eq:Winding-summation-1}
\end{align}

\section{Extended data: projected phase space trajectories and winding dynamics}

We now present extended data for several combinations of the parameters $U/J$ and $\varphi)$, showing how the addition of nonlinearities alters the projected phase space trajectories as well as the dynamics of the winding of the center of mass. Figure~\ref{FIG:s0} shows the overall trends (as a function of $U/J$ and $\varphi$) of the experimentally measured and numerically simulated average IPR and total winding, as also shown in the main text. We mainly reproduce these graphics to indicate the parameter values (indicated by pink and red squares and stars) for which we will later plot the phase space trajectories and winding dynamics in subsequent figures.

In Fig.~\ref{FIG:s1}, we plot the projected phase space dynamics of experiment and theory for $\varphi = 9\pi/16$ (pink markers), both for $U/J = 0$ (squares) and for $U/J = 2$ (stars). Figure~\ref{FIG:s2} depicts the dynamics of the winding for these same parameter values.

In Fig.~\ref{FIG:s3}, we plot the projected phase space dynamics of experiment and theory for $\varphi = 25\pi/16$ (red markers), both for $U/J = 0$ (squares) and for $U/J = 2$ (stars). Figure~\ref{FIG:s4} depicts the dynamics of the winding for these same parameter values.

\begin{figure*}[b!]
 \includegraphics[width=0.8\columnwidth]{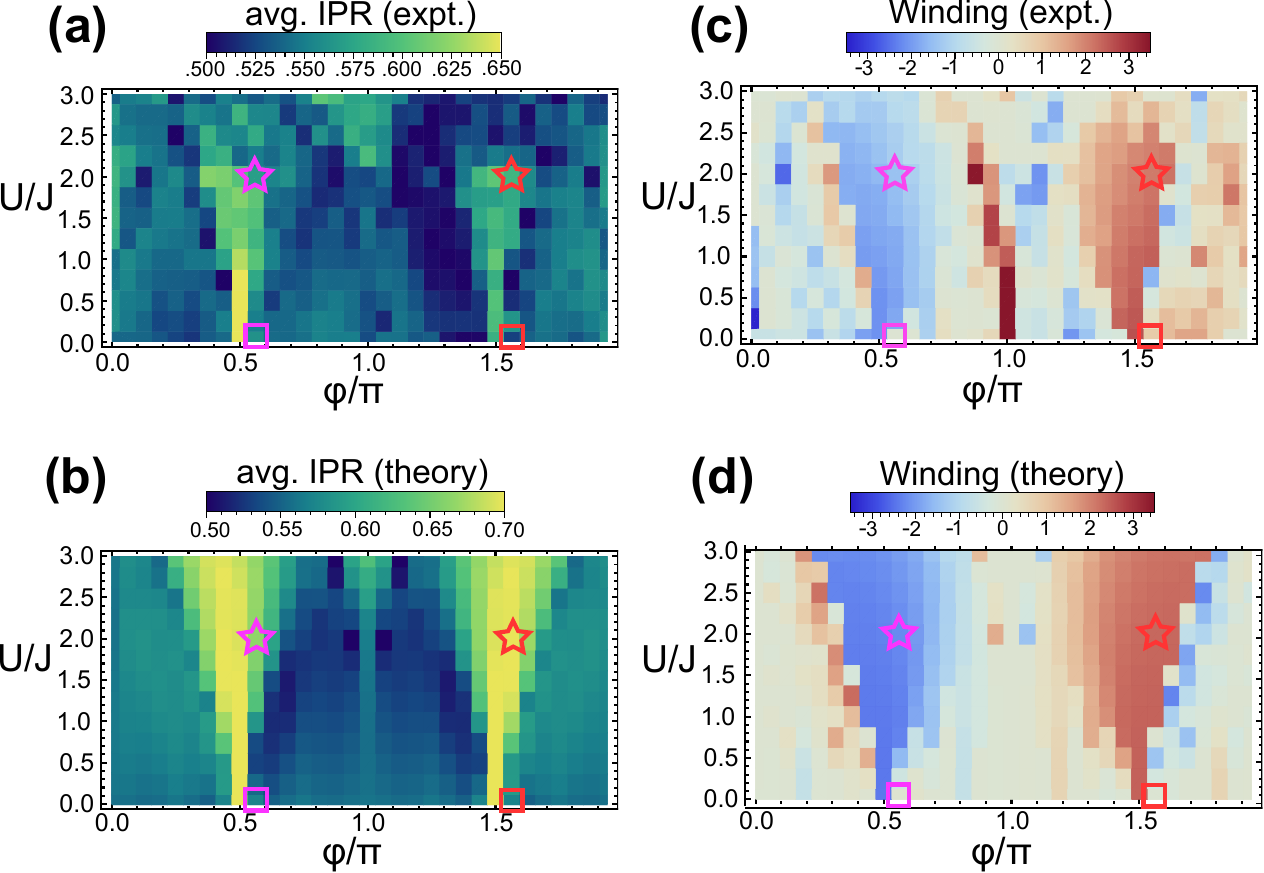}
    \centering
    \caption{\label{FIG:s0}
    \textbf{Dependence of averaged IPR and net winding on the applied flux and strength of nonlinearity.}
    The color plots of averaged IPR (a,b) and net winding (c,d) are as shown in the main text. We additionally denote selected points in parameter space for which we present extended data in subsequent supplementary figures.
    Selected points are at zero (squares) and intermediate (stars) nonlinear interaction. Pink markers are at $\frac{9\pi}{16}$ flux, red at $\frac{25\pi}{16}$. At zero nonlinear interaction, chiral behavior appears only at $\varphi = \frac{\pi}{2},\frac{3\pi}{2}$, and falls off rapidly away from that point. Adding the nonlinear interaction leads to localized, chiral dynamics away from those two points.
	}
\end{figure*}

\begin{figure*}[t!]
	\includegraphics[width=0.9\columnwidth]{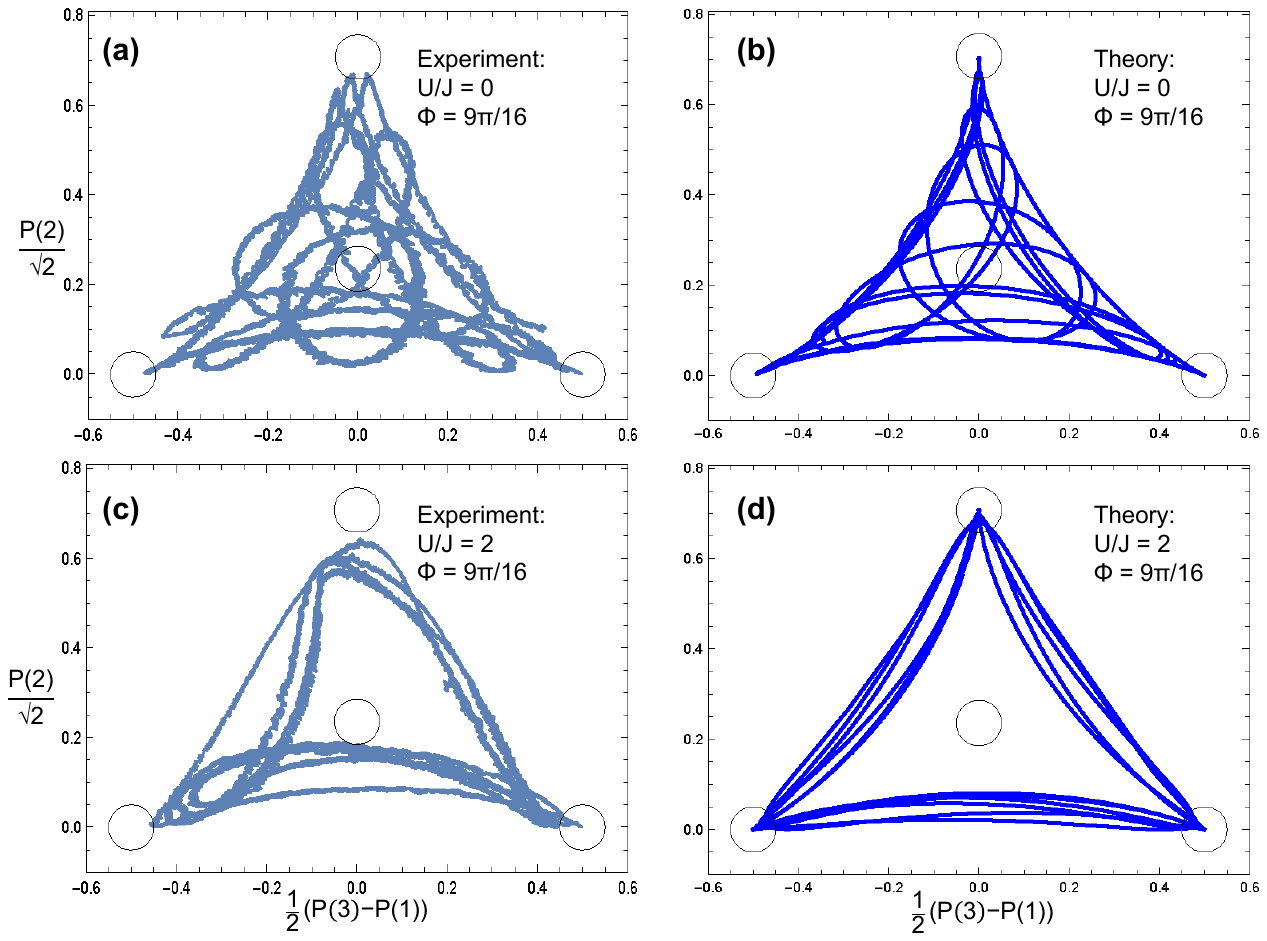}
	\centering
	\caption{\label{FIG:s1}
		\textbf{Projected phase space dynamics for a flux of $9\pi / 16$.}
	   \textbf{(a,b)}~Parametric plots of the experimentally measured and numerically simulated
    center of mass trajectory/population imbalances of the three oscillators at $\varphi = \frac{9\pi}{16}$ and no nonlinear interaction over 120 seconds of evolution time with a tunneling rate of $J/2\pi = 28$~mHz. Roughly speaking, the trajectory shows no net winding.
        \textbf{(c,d)}~Parametric plots of the experimentally measured and numerically simulated center of mass trajectory at the same flux, but with nonlinear interaction strength $U/J = 2$. The excitation winds around the triangular projected phase space, displaying winding and a relative localization.
	}
\end{figure*}

\begin{figure*}[t!]
	\includegraphics[width=0.9\columnwidth]{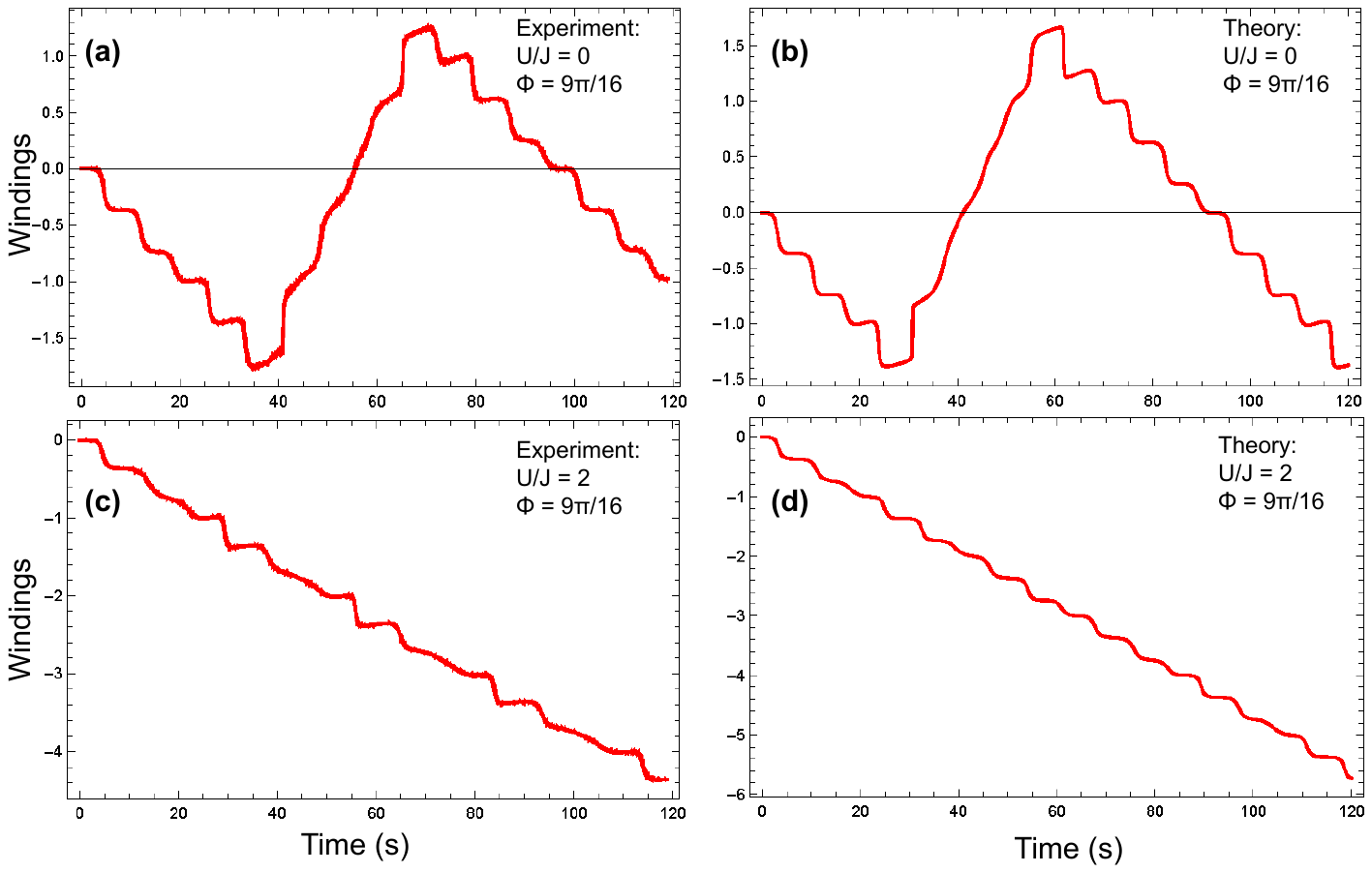}
	\centering
	\caption{\label{FIG:s2}
		\textbf{Dynamics of the winding for a flux of $9\pi / 16$.}
		\textbf{(a,b)}~Winding of the excitation around the plaquette as a function of time for $\varphi = \frac{9\pi}{16}$ and no nonlinear self-interaction. The excitation reverses course around the three sites, leading to no net winding.
        \textbf{(c,d)}~Winding for nonlinear interaction strength $U/J = 2$. The excitation travels in a single direction and does not reverse course, leading to a net chiral motion.
	}
\end{figure*}

\begin{figure*}[t!]
	\includegraphics[width=0.9\columnwidth]{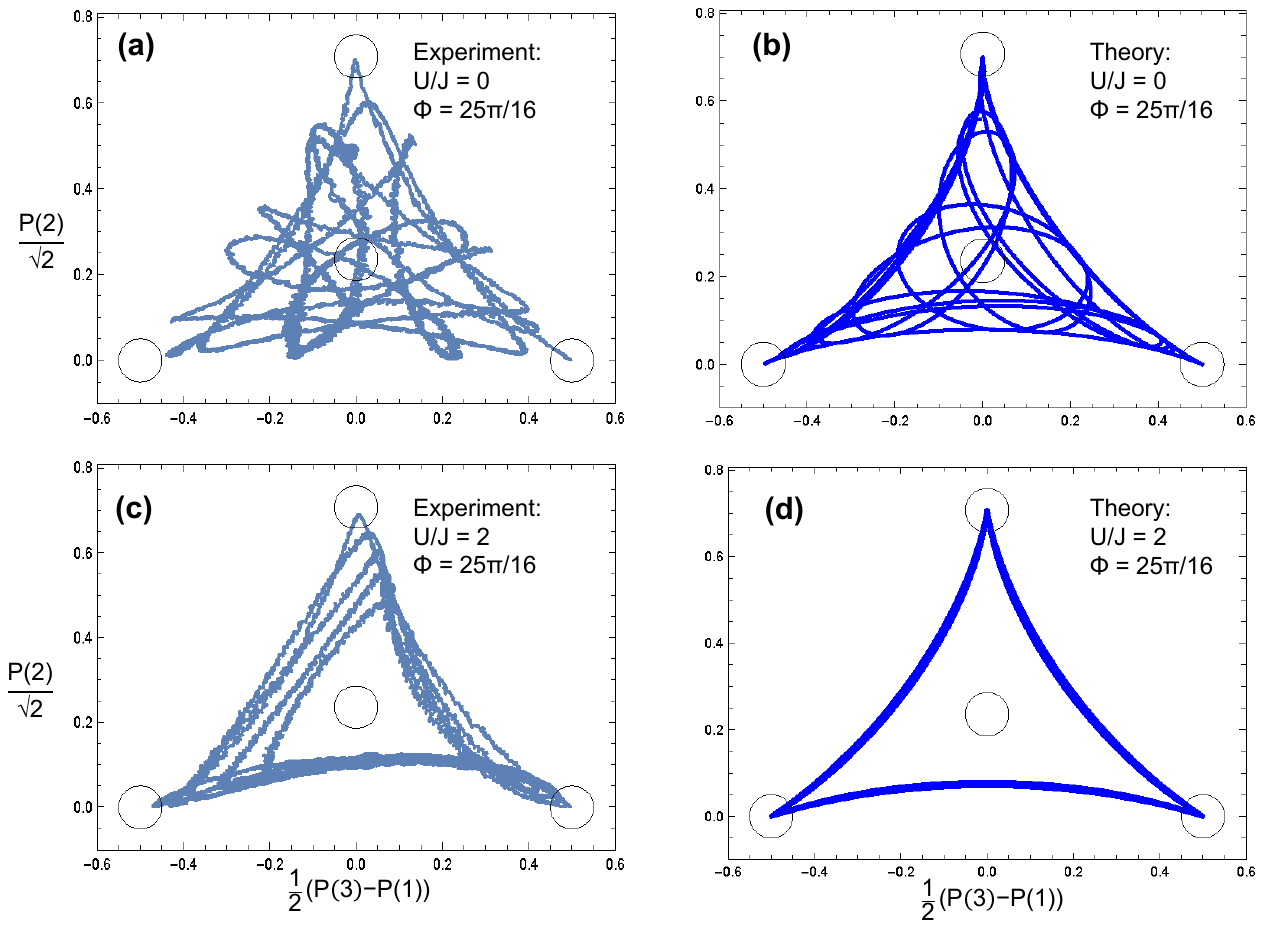}
	\centering
	\caption{\label{FIG:s3}
		\textbf{Projected phase space dynamics for a flux of $25\pi / 16$.}
		\textbf{(a,b)}~Parametric plots of the experimentally measured and numerically simulated
    center of mass trajectory/population imbalances of the three oscillators at $\varphi = \frac{25\pi}{16}$ and no nonlinear interaction over 120 seconds of evolution time with a tunneling rate of $J/2\pi = 28$~mHz. The trajectory roughly shows no net winding.
        \textbf{(c,d)}~Parametric plots of the experimentally measured and numerically simulated center of mass trajectory at the same flux, but with nonlinear interaction strength $U/J = 2$. The excitation winds around the triangular projected phase space, displaying winding and a relative localization.        
	}
\end{figure*}

\begin{figure*}[t!]
	\includegraphics[width=0.9\columnwidth]{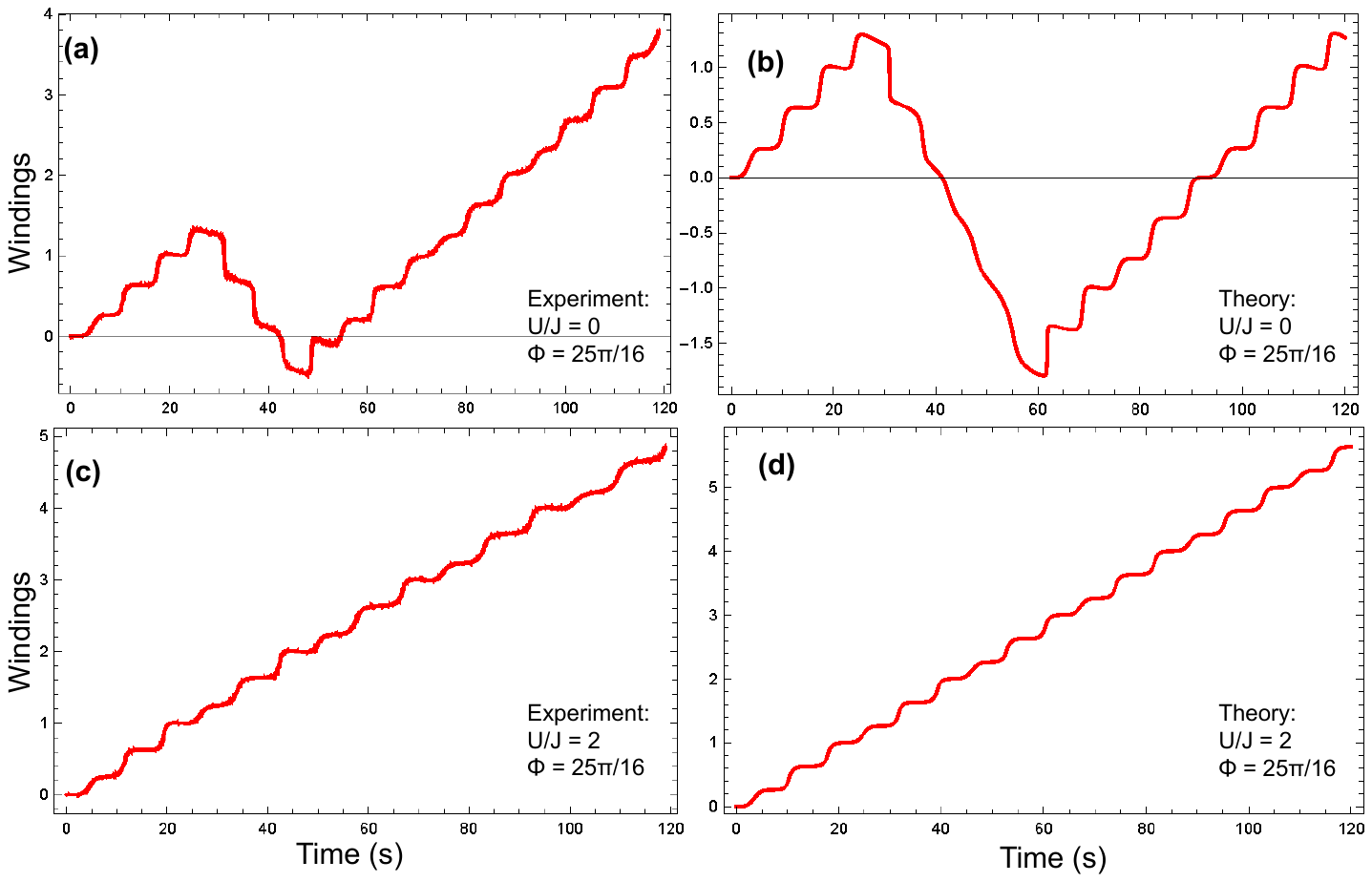}
	\centering
	\caption{\label{FIG:s4}
	\textbf{Dynamics of the winding for a flux of $25\pi / 16$.}
	\textbf{(a,b)}~Winding of the excitation around the plaquette as a function of time for $\varphi = \frac{25\pi}{16}$ and no nonlinear self-interaction. The excitation reverses course, leading to little net winding. In the experimental case, non-idealities lead to a small net motion in one direction.
    \textbf{(c,d)}~Winding for nonlinear interaction strength $U/J = 2$. The excitation travels in a single direction and does not reverse course, leading to a net chiral motion.
	}
\end{figure*}

\bibliographystyle{apsrev4-1}
\bibliography{3site}

\end{widetext}